\documentclass[11pt]{article}
\usepackage[english]{babel}
\usepackage{amsmath,amssymb,exscale,color,graphicx}
\usepackage[utf8]{inputenc}
\usepackage{amsfonts}
\usepackage{amsthm}
\usepackage{mathrsfs}
\usepackage{xcolor}
\usepackage{caption}
\usepackage{subcaption}
\usepackage{enumitem}
\usepackage{dcolumn}
\usepackage{bm}
\usepackage{multirow}
\usepackage{cite,color,url}
\usepackage[colorlinks=true
,urlcolor=blue
,anchorcolor=blue
,citecolor=blue
,filecolor=blue
,linkcolor=blue
,menucolor=blue
,linktocpage=true
,pdfproducer=medialab
,pdfa=true
]{hyperref}
\usepackage{slashed}
\usepackage{epsfig,psfrag,rotating,soul}
\usepackage{rotfloat}
\evensidemargin \oddsidemargin
\allowdisplaybreaks
\usepackage{graphicx}
\usepackage{color}
\newtheorem{theorem}{Theorem}
\newtheorem{proposition}{Proposition}
\newtheorem{lemma}{Lemma}
\begin{document}
\thispagestyle{empty}
\def\thefootnote{\fnsymbol{footnote}}

\vspace*{1cm}

\begin{center}

\begin{Large}
\textbf{\textsc{Disentangling the Seesaw in the Left-Right Model -- An Algorithm for the General Case
}}
\end{Large}

\vspace{1cm}

{\sc
Joshua~Kiers$^{1}$%
\footnote{{\tt \href{mailto:jkiers@marian.edu}{jkiers@marian.edu}}}%
, Ken~Kiers$^{2}$%
\footnote{{\tt \href{mailto:knkiers@taylor.edu}{knkiers@taylor.edu}}}%
, Alejandro~Szynkman$^{3}$%
\footnote{{\tt \href{mailto:szynkman@fisica.unlp.edu.ar}{szynkman@fisica.unlp.edu.ar}}}%
, Tatiana~Tarutina$^{4}$%
\footnote{{\tt \href{mailto:tarutina@fisica.unlp.edu.ar}{tarutina@fisica.unlp.edu.ar}}}%
}

\vspace*{.7cm}

{\sl
$^1$Department of Mathematical and Computational Sciences, Marian University, \\ 
3200 Cold Spring Rd., Indianapolis, IN 46222, United States

\vspace*{0.1cm}

$^2$Physics Department, Taylor University, \\ 
1846 Main Street, Upland, Indiana 46989, United States

\vspace*{0.1cm}

$^3$IFLP, CONICET - Dpto. de F\'{\i}sica, Universidad Nacional de La Plata, \\ 
C.C. 67, 1900 La Plata, Argentina

\vspace*{0.1cm}

$^4$IFLP, CONICET, \\ 
Diagonal 113 e/ 63 y 64, 1900 La Plata, Argentina

}

\end{center}

\vspace{0.1cm}

\date{\today}

\begin{abstract}
\noindent
Senjanovi{\'c} and Tello have analyzed how one could
determine the neutrino Dirac mass matrix in the minimal left-right model,
assuming that the mass matrices for the light and heavy neutrinos could
be taken as inputs.  They have provided an analytical
solution for the Dirac mass matrix in the case
that the left-right symmetry is implemented via a generalized
parity symmetry and that this symmetry remains unbroken 
in the Dirac Yukawa sector.  
We extend the work of Senjanovi{\'c} and Tello to the case in which the generalized
parity symmetry is broken in the Dirac Yukawa sector.
In this case the elegant method outlined by Senjanovi{\'c} and Tello
breaks down and we need to adopt a numerical approach.  Several iterative
approaches are described; these are found to work in some cases but to be highly 
unstable in others.
A stable, prescriptive numerical algorithm is described that works in all but a vanishingly
small number of cases.  We apply this algorithm to numerical data sets that
are consistent with current experimental constraints on neutrino masses 
and mixings.  We also provide some additional context and supporting explanations 
for the case in which the parity symmetry is unbroken.
\end{abstract}

\def\thefootnote{\arabic{footnote}}
\setcounter{page}{0}
\setcounter{footnote}{0}

\newpage

\section{Introduction} 
\label{sec:introduction}

The observation of neutrino oscillations~\cite{Super-Kamiokande:1998kpq} proved that at least two neutrinos are massive particles. 
However, the origin of neutrino mass is still an open question; better understanding of this fundamental
issue could yield key insights into the nature of 
physics beyond the Standard Model (SM). The seesaw mechanism is an appealing possibility 
that could account for the smallness of the neutrino mass scale~\cite{Minkowski:1977sc,Mohapatra:1979ia,Yanigida:1979dis,Glashow:1980}.
There is not one unique seesaw model for neutrino mass, however, so even confirming that the 
seesaw mechanism is the source of neutrino mass does not necessarily lead to a complete
understanding of the underlying model.

Seesaw models generically contain two types of Yukawa terms that couple the Higgs and lepton fields.
The first is familiar from the SM and couples the left- and right-handed projections 
of the lepton fields.  In the neutral sector, the 
resulting mass matrix is called the Dirac mass matrix and is denoted by $M_D$.
The second couples the left-handed projections of the lepton fields
to their charge-conjugates, and similarly for the right-handed fields,
giving rise to so-called Majorana mass terms.  In the seesaw mechanism, the mass matrix for the light neutrinos
results from the interplay between these Dirac and Majorana mass terms.

As noted in Ref.~\cite{Senjanovic:2018xtu}, it is interesting to compare the situation for neutrinos
to that for the charged fermions.  In the Standard Model, the charged fermions
receive their masses through their Yukawa interactions with the Higgs field.
As a result, the measured values of their masses lead directly to
predictions for the partial widths for Higgs decays into fermion-antifermion pairs.
Measurements made at the Large Hadron Collider (LHC) have so far been compatible with the
SM predictions~\cite{Workman:2022ynf}.
The situation is considerably more complicated in the neutrino sector 
if the seesaw mechanism is at play.  
In this case, following the analogy from the charged fermions, one would 
want to determine the elements of the Dirac mass matrix $M_D$
as a function of the light neutrino masses and mixings 
and those of the heavy states.\footnote{A reverse path is adopted in Refs.~\cite{Falcone:2003af,Akhmedov:2005np,Hosteins:2006ja,Akhmedov:2006de} where the
heavy neutrino mass matrix is determined from $M_D$ and the light
neutrino mass matrix.}  The former are being carefully investigated
at current neutrino experiments and the latter could possibly be measured at the LHC or 
some future collider~\cite{Senjanovic:2018xtu}.

In the simplest formulation of the seesaw mechanism,
the Dirac mass matrix $M_D$ cannot be uniquely determined from the light and heavy mass matrices 
($M_\nu$ and $M_N$, respectively), since it can always be redefined by an arbitrary 
complex orthogonal matrix~\cite{Casas:2001sr,deGouvea:2010iv}.  Within the context of the left-right 
symmetric extension of the SM~\cite{Pati:1974yy,Mohapatra:1974gc,Senjanovic:1975rk,Senjanovic:1978ev}, however, where the seesaw mechanism 
arises as a direct consequence of spontaneous left-right symmetry breaking,
the matrix $M_D$ is defined only in terms of physical quantities
(see Refs.~\cite{Nemevsek:2012iq,Senjanovic:2016vxw,Senjanovic:2018xtu,Senjanovic:2019moe}).
At the level of the underlying model, left-right symmetry may be implemented by imposing a
generalized charge conjugation symmetry, ${\cal C}$, or a generalized parity symmetry, ${\cal P}$.
The charge conjugation approach has been analyzed in Ref.~\cite{Nemevsek:2012iq}; in this case
it is argued that the relation between
$M_D$ and the masses and mixings of the light and heavy neutrino states is significantly
simplified due to the fact that $M_D$ is symmetric.
As noted in Ref.~\cite{Senjanovic:2018xtu}, however, the 
generalized parity case is highly nontrivial and contains two distinct possibilities,
depending on whether or not ${\cal P}$ remains unbroken in the Dirac Yukawa sector.
If ${\cal P}$ is unbroken in the Dirac Yukawa sector, $M_D$ is Hermitian\footnote{Actually, $M_D$
is Hermitian up to multiplication by a diagonal sign matrix in this case.  Please see below for further details.}
and may be determined analytically, given
the masses and mixings of the light and heavy neutrino states~\cite{Senjanovic:2018xtu, Senjanovic:2019moe}.
We shall refer to this as the ``parity-conserving'' scenario in the remainder of this
work.  By way of contrast, in the ``parity-violating'' scenario $M_D$ is no longer 
Hermitian\footnote{To be more precise,
in this case the Vacuum Expectation Values (VEVs) of the bidoublet Higgs field contain a CP-violating phase that breaks the generalized
parity symmetry in the Dirac Yukawa sector and leads to $M_D$ no longer being Hermitian.}
and the analytical approach to determining $M_D$ breaks down.  Nevertheless,
the authors of Ref.~\cite{Senjanovic:2018xtu} develop a phenomenological analysis 
that would possibly allow one to determine $M_D$ through the study of specific processes in this scenario.
In summary, the left-right model, in which the parity-violating nature of the weak interactions 
follows from the spontaneous breaking of the left-right symmetry, 
turns out to be a theoretical picture that not only results in non-zero neutrino mass but 
also elucidates its origin.

In this article we develop a general method to determine $M_D$ in the left-right model in the case
that the underlying model (before spontaneous symmetry breaking) is invariant under ${\cal P}$.  
Our determination of $M_D$ only relies 
on the knowledge of $M_\nu$ and $M_N$, as well as information related to the VEVs of the bidoublet Higgs field.  
In this way it is analogous to the analytical solution found 
in Refs.~\cite{Senjanovic:2016vxw,Senjanovic:2018xtu,Senjanovic:2019moe} for the parity-conserving limit, where $M_D$ is obtained directly
from these two matrices and the ratio of the bidoublet Higgs VEVs.  What distinguishes our approach from previous
work is that our approach does not 
assume that parity is conserved in the Dirac Yukawa sector after spontaneous symmetry breaking.
That is, we assume that $M_D$ could be non-Hermitian.
In previous approaches, determination of $M_D$ in the parity-violating case required
the study of additional specific processes.\footnote{As 
noted above, the method proposed in this paper requires the masses and mixings of the heavy neutrinos contained 
in $M_N$, therefore those processes~\cite{PhysRevLett.50.1427,Das:2012ii,Aguilar-Saavedra:2012grq,Vasquez:2014mxa,Gluza:2016qqv,Dev:2015kca,Das:2017hmg} that would allow one to determine 
that information are still necessary.}

To determine $M_D$ from $M_\nu$ and $M_N$ one needs to solve a system of non-linear matrix equations. 
The solution presented for the parity-conserving case in Refs.~\cite{Senjanovic:2016vxw,Senjanovic:2018xtu,Senjanovic:2019moe} is obtained through 
an ingenious procedure that makes use of the hermiticity of $M_D$.  When parity is broken,
$M_D$ may no longer be assumed to be Hermitian and the procedure breaks down. In this case, the authors show 
that a solution can in principle be obtained numerically as an expansion in a small parameter, 
although they do not present a specific numerical algorithm to implement this strategy.
A key feature of their proposed strategy in the parity-violating case is that the non-Hermitian matrix $M_D$ is replaced
by a Hermitian matrix that is a function of $M_D$. The method presented in the current work extends the approach
outlined in Ref.~\cite{Senjanovic:2018xtu}
and provides a systematic prescription to solve for $M_D$ numerically, even in the case that it
is non-Hermitian. In a sense one could say that it adds a 
piece to the puzzle of unwinding the seesaw as the origin of neutrino mass. 
In the parity-conserving case, the analytical solution
described in Refs.~\cite{Senjanovic:2016vxw,Senjanovic:2018xtu,Senjanovic:2019moe} allows one to resolve the seesaw by determining
$M_D$ in terms of $M_\nu$ and $M_N$.
In the parity-violating case one could use the phenomenological approach outlined in Refs.~\cite{Senjanovic:2016vxw,Senjanovic:2018xtu} to 
determine $M_D$.  Of course, one could also apply that phenomenological 
analysis when parity is conserved, which would allow one to cross-check the values of $M_D$ extracted from experiment with those 
corresponding to the analytical solution. Likewise, in the parity-violating case, the matrix $M_D$ that would 
result from experiment could also be compared with the method proposed in this study, allowing one to further resolve the puzzle.

To illustrate our method and study its performance we undertake a numerical analysis using several theoretical
data sets that are compatible with the
current experimental constraints on the lepton masses and mixings.  We use a Monte Carlo algorithm to generate these data sets, 
following the framework and approach 
described in Refs.~\cite{Kiers:2005vx,Kiers:2002cz}.
The Monte Carlo algorithm provides $M_\nu$, $M_N$ and $M_D$ in each case, allowing us to 
take the matrices $M_\nu$ and $M_N$ as inputs 
and to determine whether our approach is able to recover the corresponding matrix $M_D$.
Most of the data sets are of the parity-violating 
variety, but we also consider one parity-even data set so that we can test our method in this case as well.
We find that our method successfully obtains a solution for $M_D$ for all of the data sets that we study.
Throughout this analysis we assume the normal ordering of neutrino 
masses\footnote{This is an arbitrary choice; an inverted ordering could be considered as well.} and 
ignore the possibility of light sterile neutrinos.  

Our paper is organized as follows. In Section~\ref{sec:model} we outline the specific version of the
Left-Right Model that we employ, including various details about our notation.
Section~\ref{sec:method-description} contains a detailed description of our method for determining $M_D$.
This method is then illustrated with three numerical examples in 
Section~\ref{sec:results}.  In Section~\ref{sec:other-methods} we briefly discuss alternative methods that 
we had previously used in our attempts to determine a solution for $M_D$.  These approaches
were successful for some of the data sets, but were unstable for others, illustrating the significant challenge posed by solving
the set of nonlinear matrix equations to determine $M_D$.
Section~\ref{sec:derivations} contains a proof of a key mathematical relation used in Section~\ref{sec:method-description},
as well as derivations of several mathematical properties for the parity-conserving case considered
in Refs.~\cite{Senjanovic:2018xtu, Senjanovic:2019moe}.
We conclude with a brief discussion of our results
in Section~\ref{sec:discussion-conclusions}. Finally, the Appendices outline the diagonalization of the charged and neutral mass matrices, details of the notation and specifics about the method and also include some mathematical results related to the parameterization of complex orthogonal matrices.

\section{The Model}
\label{sec:model}

In this section we provide a brief summary of the Left Right Model (LRM),
primarily following the notation and conventions used in Ref.~\cite{Kiers:2005vx};
the interested reader is referred to Ref.~\cite{Kiers:2005vx} for more
detail.

The underlying symmetry of the LRM is based on the 
gauge group $SU(2)_L\times SU(2)_R\times U(1)_{B-L}$.
The specific formulation of the LRM considered in 
Ref.~\cite{Kiers:2005vx} contains two Higgs triplet fields,
\begin{eqnarray}
	\Delta_{L,R} = \left(\begin{array}{cc}
		\delta^+_{L,R}/\sqrt{2} & \delta^{++}_{L,R} \\
		\delta^0_{L,R} & -\delta^+_{L,R}/\sqrt{2} \\
		\end{array}\right) \; ,
\end{eqnarray}
as well as a bidoublet Higgs field,
\begin{eqnarray}
	\phi = \left(\begin{array}{cc}
		\phi_1^0 & \phi_1^+ \\
		\phi_2^- & \phi_2^0 \\
		\end{array}\right)\; .
\end{eqnarray}
The Yukawa terms for the charged and neutral leptons
may then be written as~\cite{Kiers:2005vx}
\begin{eqnarray}
	-{\mathcal L}_\textrm{\scriptsize Yukawa} = 
		\overline{\psi}_{iL}^\prime\left(
		G_{ij} \phi + H_{ij} 
		\widetilde{\phi}\right)\psi_{jR}^\prime +
		\frac{i}{2}F_{ij}\left(\psi^{\prime T}_{iL} C \tau_2
		\Delta_L\psi^{\prime}_{jL} +
		\psi^{\prime T}_{iR} C \tau_2
		\Delta_R\psi^{\prime}_{jR}\right)
		+ \textrm{h.c.} \; ,
	\label{eq:yuk}
\end{eqnarray}
where $C=i\gamma^2\gamma^0$ and $\widetilde{\phi}= \tau_2 \phi^* \tau_2$,
and where $\psi_{iL,R}^\prime$ represent the left- and right-handed
lepton doublets in the gauge basis,
\begin{eqnarray}
	\psi^\prime_{iL,R} = \left(\begin{array}{c}
		\nu^\prime_{iL,R}\\
		e^\prime_{iL,R}\\
		\end{array}\right) \;,
\end{eqnarray}	
where $i$ is a generation index.  The matrices $G$ and $H$ 
are taken to be Hermitian, while $F$ may be assumed to be complex symmetric without loss of 
generality.\footnote{See the discussion in Ref.~\cite{Kiers:2005vx}, as well 
as Refs.~\cite{Deshpande:1990ip, Kayser:2002qs}.}  The model also contains an extra left-right
parity symmetry, ${\cal P}$~\cite{Senjanovic:2018xtu, Deshpande:1990ip, Kiers:2005vx}, under which
\begin{eqnarray}
	\psi^\prime_{iL}\leftrightarrow\psi^\prime_{iR},
	~~~\phi\leftrightarrow \phi^\dagger,
	~~~\Delta_L \leftrightarrow \Delta_R.
	\label{eq:parity}
\end{eqnarray}

The neutral Higgs fields obtain VEVs upon spontaneous
symmetry breaking; the Higgs VEVs may be parameterized as 
follows,\footnote{Reference~\cite{Kiers:2005vx} uses the phase $\alpha = \pi -a$; 
here we follow the notation of Ref.~\cite{Senjanovic:2018xtu} for the phase.}
\begin{eqnarray}
	\langle \phi\rangle =\left(\begin{array}{cc}
		k_1/\sqrt{2} & 0 \\
		0& -k_2e^{-ia}/\sqrt{2} \\
		\end{array}\right) ,~~~
	\langle \Delta_L \rangle= \left(\begin{array}{cc}
		0 & 0 \\
		v_Le^{i\theta_L}/\sqrt{2} & 0\\
		\end{array}\right),~~~
	\langle \Delta_R \rangle= \left(\begin{array}{cc}
		0 & 0 \\
		v_R/\sqrt{2} & 0\\
		\end{array}\right),
	\label{eq:vevs}
\end{eqnarray}
where $k_1$, $k_2$, $v_L$ and $v_R$ are all taken to be real and positive.
If the phase $a$ is a multiple of $\pi$, then $\langle \phi\rangle$
respects the generalized parity symmetry even after spontaneous symmetry breaking; 
we refer to this as the ``parity-conserving'' case insofar as the Dirac Yukawa sector is concerned.
Experimental constraints suggest
$v_R\gg k_1, k_2\gg v_L$; also, we have~\cite{Langacker:1989xa}
\begin{eqnarray}
	k_1^2+k_2^2 \simeq 
		\frac{4m_W^2}{g^2}\simeq(246.2~\textrm{GeV})^2 \; .
		\label{eq:kkpr1}
\end{eqnarray}
As noted in Ref.~\cite{Kiers:2005vx}, it is natural to assume that the 
ratio $k_2/k_1$ is of order $m_b/m_t$.

The Yukawa terms in the Lagrangian lead to mass terms for the charged and neutral 
leptons when the neutral Higgs fields acquire VEVs.
The mass matrix for the charged leptons in the gauge basis is given by~Ref.~\cite{Kiers:2005vx}
\begin{eqnarray}
	M_\ell = \frac{1}{\sqrt{2}}\left(-Gk_2e^{-ia}+H k_1 \right)\; .
		\label{eq:mlep}
\end{eqnarray}
Recalling that $G$ and $H$ are Hermitian, we see that $M_\ell$ is Hermitian
if the phase $a$ is a multiple of $\pi$ (i.e., in the so-called parity-conserving case).

The neutral lepton sector is more complicated than the charged lepton sector, 
since the Yukawa Lagrangian generically leads to Majorana mass terms
in addition to the ``ordinary'' Dirac mass terms.  For three lepton generations, the mass matrix
is a $6\times 6$ complex symmetric matrix,
\begin{eqnarray}
	\left(\begin{array}{cc}
			M_{LL}^\dagger & M_{LR}\\
			M_{LR}^T & M_{RR}\\
			\end{array}\right) \; ,
		\label{eq:6by6}
\end{eqnarray}
where
\begin{eqnarray}
	M_{LR} & = & \frac{1}{\sqrt{2}}\left(Gk_1-H k_2e^{ia}\right)\;
		\label{eq:MLR}
\end{eqnarray}
is a $3\times 3$ Dirac mass matrix and where
\begin{eqnarray}
	M_{LL} & = & \frac{1}{\sqrt{2}}F v_Le^{i\theta_L}
		\label{eq:MLL}
\end{eqnarray}
and
\begin{eqnarray}
	M_{RR} = \frac{1}{\sqrt{2}}F v_R \; ,
    \label{eq:MRR}
\end{eqnarray}
are $3\times 3$ Majorana mass matrices for the left- and right-handed fields, respectively.
As is evident from the above expressions, $M_{LL}$ and $M_{RR}$ are both complex
symmetric matrices; $M_{LR}$ is Hermitian if the phase $a$ is a multiple of $\pi$.
Equation~(\ref{eq:6by6}) may be approximately block diagonalized (see Ref.~\cite{Kiers:2005vx} for details),
which leads to $3\times 3$ complex symmetric mass matrices for the (mostly) left- and 
(mostly) right-handed fields.  The mass matrix for the right-handed neutrinos is simply $M_{RR}$, so that
the right-handed neutrinos are generically quite heavy (due to the assumed large value of $v_R$).
The mass matrix for the left-handed neutrinos is
\begin{eqnarray}
	M_{LL}^\dagger-M_{LR}M_{RR}^{-1}M_{LR}^T \; .
    \label{eq:Mnu-2005}
\end{eqnarray}
The first term is small, since it is proportional to $v_{L}$.
The second term is suppressed due to the presence of $M_{RR}^{-1}$; this
suppression is known as the seesaw mechanism.

We have so far been working in the gauge basis.
To make connections to measurable quantities, one needs to diagonalize 
the mass matrices for the charged and neutral leptons, which yields
the physical lepton masses, as well as the Pontecorvo-Maki-Nakagawa-Sakata (PMNS) matrix.  These
quantities may then be measured
or constrained by various types of experiments.
The basis in which the mass matrices for the charged and neutral leptons
are all diagonal is called the mass basis.

In the remainder of this paper we adopt the notation used by
Senjanovi\'c et al in Ref.~\cite{Senjanovic:2018xtu} and work in a basis that is part-way between the gauge
basis and the mass basis~\cite{Senjanovic:2018xtu}.  In this basis,
which we refer to as the ``charged-diagonal'' basis, one diagonalizes the mass matrix for
the charged leptons and implements a corresponding transformation on
the neutrino mass matrices.  The neutrino mass matrices are not generally diagonal
in this basis.

Table~\ref{table:mass-matrices} shows the correspondence between the 
mass matrices in the charged-diagonal basis (of Senjanovi\'c et al)
and those in the gauge basis (described above and in Ref.~\cite{Kiers:2005vx}).
The table does not explicitly include
the unitary matrices that are necessary to go from the gauge basis to the
charged-diagonal basis.
The interested reader is referred to Appendix~\ref{app:model}
for the precise relations between quantities in these two bases.

\begin{table}
\begin{center}
\begin{tabular}{ccc}
    Senjanovi\'c, et al (Ref.~\cite{Senjanovic:2018xtu}) and present work & & Kiers, et al (Ref.~\cite{Kiers:2005vx}) \\
    \hline
    $m_e$ (diagonal) & $\leftrightarrow$ & $M_\ell$ \\
    $M_D$ & $\leftrightarrow$ & $M_{LR}^\dagger$ \\
    $M_N$ & $\leftrightarrow$ & $M_{RR}^*$ \\
    $M_\nu$ & $\leftrightarrow$ & $\left(M_{LL}^\dagger-M_{LR}M_{RR}^{-1}M_{LR}^T\right)^*$ \\
\end{tabular}
\end{center}
\caption{Correspondence between various mass matrices in Senjanovi\'c, et al (Ref.~\cite{Senjanovic:2018xtu})
and those in Kiers, et al (Ref.~\cite{Kiers:2005vx}).  Senjanovi\'c et al work primarily in a basis in which
the charged lepton mass matrix is diagonal, whereas Kiers et al work in the gauge basis, in which neither
the charged nor the neutral lepton mass matrices are assumed to be diagonal.  The precise 
relations between the various matrices is given in Appendix~\ref{app:model}.}
\label{table:mass-matrices}
\end{table}

It is straightforward to derive the following three relations between various mass
matrices in the charged-diagonal basis,
\begin{equation}
	M_D-U_eM_D^\dagger U_e = is_at_{2\beta}(e^{ia}t_\beta M_D +m_e) \, ,
	\label{eq26}
\end{equation}
\begin{equation}
	U_em_eU_e-m_e = is_at_{2\beta}(M_D+e^{-ia}t_\beta m_e) \, ,
	\label{eq27}
\end{equation}
\begin{equation}
	M_{\nu} = \frac{v_L e^{i\theta_L}}{v_R}U_e^T M_N^* U_e - M_D^T\frac{1}{M_N}M_D \, ,
	\label{eq31}
\end{equation}
where $\tan\beta \equiv k_2/k_1$ and where
the unitary matrix $U_e$ is associated with the transformation from
the gauge basis to the charged-diagonal basis 
[see Eq.~(\ref{eq:Ue-def}) in the Appendix for a precise definition].
Also, $s_a$, $t_\beta$ and $t_{2\beta}$ stand for $\sin(a)$, 
$\tan\beta$ and $\tan\left(2\beta\right)$, respectively.
Note that there is an overall sign ambiguity for $U_e$ in the sense that Eqs.~(\ref{eq26})-(\ref{eq31})
are unchanged under $U_e \to -U_e$.

\section{Description of Method}
\label{sec:method-description}

Our primary goal in this work is to describe a method that can be used
to solve Eqs.~(\ref{eq26}), (\ref{eq27}) and (\ref{eq31}) for $U_e$ and $M_D$,
taking $M_N$, $M_\nu$, $v_l e^{i \theta_L}/v_R$, $a$ and $\beta$ as inputs.  
The authors of 
Ref.~\cite{Senjanovic:2018xtu} outlined such a procedure in the case that $s_at_{2\beta} = 0$.
In that case, Eqs.~(\ref{eq26}) and (\ref{eq27}) reduce to the
expressions $M_D = U_e M_D^\dagger U_e$ and $m_e = U_e m_e U_e$, respectively.  
Recalling that $m_e$ is a real, diagonal matrix, we see that
$U_e = \mbox{diag}(\pm 1, \pm 1, \pm 1)$ in this case,
and that $M_D U_e^\dagger$ is Hermitian (we say that $M_D$ is ``sign Hermitian'').  
Armed with this knowledge, Senjanovi\'c et al 
were able to work out an analytical method to solve for $M_D$.
They were also able to classify the solutions into various
categories.

When $s_at_{2\beta} \neq 0$, the analytical method devised in Ref.~\cite{Senjanovic:2018xtu} breaks down,
since $M_D$ is no longer sign Hermitian.  As we shall see, in this case it is possible
to find a solution of Eqs.~(\ref{eq26}), (\ref{eq27}) and (\ref{eq31}) by using
an iterative approach.  We have applied this approach to various data sets and it appears
to be quite stable (see Sec.~\ref{sec:results} for further details).\footnote{We have also devised a
number of other iterative approaches that are stable for some data sets, but not for others.  These
are described in Sec.~\ref{sec:other-methods}.}

The starting point for our method is to define the matrix ${\cal M}$ 
as in Ref.~\cite{Senjanovic:2018xtu}:
\begin{equation}
	{\cal M} = \left(M_D + e^{-i a}t_\beta m_e \right) U_e^\dagger \; .
	\label{eq:calM}
\end{equation}
The algorithm described below is designed to determine ${\cal M}$, which allows one to
determine $U_e$ (see Appendix~\ref{app:Ue-determination}) and then finally to
calculate $M_D$ via
\begin{equation}
	M_D = {\cal M}U_e - e^{-i a}t_\beta m_e \; .
	\label{eq:MD}
\end{equation}
A convenient property of ${\cal M}$ is that it is Hermitian.\footnote{This follows from Eqs.~(\ref{eq26}) and (\ref{eq27}).}
Inspired by the mathematical manipulations that led to Eqs.~(40) and (41) in Ref.~\cite{Senjanovic:2018xtu}, 
we substitute Eq.~(\ref{eq:MD}) into the complex conjugate of Eq.~(\ref{eq31}), 
multiply from the left by $\frac{1}{\sqrt{M_N}}U_e$ and from the right by $U_e^T \frac{1}{\sqrt{M_N}}$,
and simplify, which yields
\begin{eqnarray}
	& &\frac{v_L e^{-i \theta_L}}{v_R} I - \frac{1}{\sqrt{M_N}} U_e M_\nu^* U_e^T \frac{1}{\sqrt{M_N}} \nonumber\\
	& & ~~~~~~~~~ =
		\frac{1}{\sqrt{M_N}}\left({\cal M} -e^{ia}t_\beta U_e m_e\right) \frac{1}{M_N^*}
		\left({\cal M}^* -e^{ia} t_\beta m_e U_e^T\right)\frac{1}{\sqrt{M_N}} \; .
\end{eqnarray}
We then define $H$, $B$ and $\tilde{H}$ as follows
\begin{eqnarray}
	H & = & \frac{1}{\sqrt{M_N}} {\cal M}\frac{1}{\sqrt{M_N^*}} 
	\label{eq:H}\\
	B & = & e^{ia}t_\beta \frac{1}{\sqrt{M_N}} U_e m_e \frac{1}{\sqrt{M_N^*}} ,
	\label{eq:B}\\
	\tilde{H} & = & H- B \; ,
	\label{eq:HtildeHB}
\end{eqnarray}
so that
\begin{eqnarray}
	\tilde{H}\tilde{H}^T = \left(H-B\right)\left(H-B\right)^T = S \; ,
	\label{eq:HtildeHtildeT}
\end{eqnarray}
where
\begin{eqnarray}
	S \equiv \frac{v_L e^{-i \theta_L}}{v_R} I - \frac{1}{\sqrt{M_N}} U_e M_\nu^* U_e^T \frac{1}{\sqrt{M_N}} .
	\label{eq:Sdefn}
\end{eqnarray}
Note that $H$ is Hermitian, since ${\cal M}$ is Hermitian and ${M_N}$ is complex symmetric.
Also, $S$ is complex symmetric.
The reader may note a certain amount of similarity
between Eqs.~(\ref{eq:H})-(\ref{eq:HtildeHtildeT}) (above) and Eqs.~(40) and (41) in Ref.~\cite{Senjanovic:2018xtu}.  
A crucial difference, however, is that $\tilde{H}$ is not generally Hermitian.

Since $S$ is symmetric, we may write Eq.~(\ref{eq:HtildeHtildeT}) in ``symmetric normal form'' as follows \cite{Gantmacher1960},
\begin{eqnarray}
	\tilde{H}\tilde{H}^T = S = O s O^T \; ,
	\label{eq:OsOT}
\end{eqnarray}
where $O$ is a complex orthogonal matrix.  In principle, the matrix $s$ could be block diagonal; in practice, we
assume that it is diagonal.  Note that the elements in $s$ are typically complex.  As shown in Section~\ref{sec:derivations},
$\tilde{H}$ itself can be written as
\begin{eqnarray}
	\tilde{H} = O \sqrt{s} \tilde{E} O^\dagger \; ,
	\label{eq:Htilde}
\end{eqnarray}
where $\tilde{E}$ is a complex orthogonal matrix.  Since $H$ is Hermitian, we may write
\begin{eqnarray}
	H - H^\dagger = 0 = B - B^\dagger + O \sqrt{s} \tilde{E} O^\dagger - O \tilde{E}^\dagger \sqrt{s}^* O^\dagger \;,
\end{eqnarray}
which may be rearranged to give
\begin{eqnarray}
	\tilde{E} = \frac{1}{\sqrt{s}}O^T\left(B^\dagger - B \right)O^* + \frac{1}{\sqrt{s}}\tilde{E}^\dagger\sqrt{s}^* \; .
	\label{eq:Etilde}
\end{eqnarray}
Denoting the $j$th element of the diagonal matrix $\sqrt{s}$ by $\sqrt{\left|s_j\right|}e^{i\gamma_j}$, where
$\gamma_j$ is taken to be real, and defining
\begin{eqnarray}
	\Delta \tilde{B} & \equiv & \frac{1}{\sqrt{s}} O^T\left(B^\dagger - B \right)O^* \; ,
	\label{eq:DeltaB}
\end{eqnarray}
we see that
\begin{eqnarray}
	\tilde{E}_{ij} = \Delta \tilde{B}_{ij} + \tilde{E}^*_{ji}\sqrt{\frac{\left|s_j\right|}{\left|s_i\right|}}e^{-i\left(\gamma_i+\gamma_j\right)} \; ,
	\label{eq:Etildeij}
\end{eqnarray}
in which there is no implied sum over repeated indices.
For future reference, we also define
\begin{eqnarray}
	\Delta_{ij} \equiv \tilde{E}_{ij} - \Delta \tilde{B}_{ij} 
	-\tilde{E}^*_{ji}\sqrt{\frac{\left|s_j\right|}{\left|s_i\right|}}e^{-i\left(\gamma_i+\gamma_j\right)} \; .
	\label{eq:Deltaij}
\end{eqnarray}
According to Eq.~(\ref{eq:Etildeij}), of course, this quantity should be zero for all $i$ and $j$.
Our numerical procedure seeks to determine values for the $\tilde{E}_{ij}$ such that
the $\Delta_{ij}$ are zero, or very close to zero.  
Once $\tilde{E}$ is determined, one can eventually compute $U_e$ and $M_D$.

\subsection{Iterative procedure to solve for $\mathbf{U_e}$ and $\mathbf{M_D}$}
\label{subsec:iterative-method}

We now outline an iterative procedure that may be used to solve for $U_e$ and $M_D$.  Except for
certain edge cases (to be described below), this procedure appears to be quite robust.
A key to the success of this algorithm is the fact that, in practice, Eq.~(\ref{eq:Sdefn}) 
is relatively well-approximated by making the replacement
$U_e\to \tilde{I}$, where $\tilde{I}$ is a diagonal matrix with $\pm 1$ (as appropriate) down the diagonal.

In the description of the algorithm that follows, we denote quantities evaluated
during the $n$-th iteration of the algorithm with an ``$n$'' subscript.  In the first step of the
$n$-th iteration of the algorithm, for example, we insert $M_\nu, M_N, a, \beta$ and $U_{e, n}$
into Eqs.~(\ref{eq:B}) and (\ref{eq:Sdefn}) and use those expressions
to calculate $B_n$ and $S_n$.  Our shorthand notation for this is as follows: 
``Eqs.~(\ref{eq:B}) and (\ref{eq:Sdefn}): $M_\nu, M_N, a, \beta, U_{e, n} \rightarrow B_n, S_n$.''
Adopting this shorthand throughout, we summarize the algorithm as follows,
\begin{enumerate}
\item Eqs.~(\ref{eq:B}) and (\ref{eq:Sdefn}): $M_\nu, M_N, a, \beta, U_{e, n} \rightarrow B_n, S_n$.
\label{step:BnSn}
\item Eq.~(\ref{eq:OsOT}): $S_n \rightarrow O_n, s_n$.
\label{step:OsOT}
\item Eq.~(\ref{eq:DeltaB}): $B_n, O_n, s_n \rightarrow \Delta \tilde{B}_n$.
\item Eq.~(\ref{eq:Etildeij}): $\Delta \tilde{B}_n, s_n \rightarrow \tilde{E}_n$;
if no solution is found for $\tilde{E}$, revise initial guess for $U_e$ (denoted $\tilde{I}$) and/or the sign
of the determinant of $\tilde{E}$, return to Step~\ref{step:BnSn} and start over.
\label{step:Etilde}
\item Eq.~(\ref{eq:Htilde}): $O_n, s_n, \tilde{E}_n \rightarrow \tilde{H}_n$.
\item Eq.~(\ref{eq:HtildeHB}): $\tilde{H}_n, B_n \rightarrow H_n$.
\item Eq.~(\ref{eq:H}): $M_N, H_n \rightarrow {\cal M}_n$.
\label{step:calM}
\item Eqs.~(\ref{eq:Ue-product})-(\ref{eq:alphai2}) (and further discussion in Appendix~\ref{app:Ue-determination}): $ m_e, \tilde{I}, a, \beta, {\cal M}_n \rightarrow U_{e, n+1}$.
\label{step:Ue}
\item Eq.~(\ref{eq:MD}): $m_e, a, \beta, {\cal M}_n, U_{e, n+1} \rightarrow M_{D,n+1}$; return to Step~\ref{step:BnSn}.
\label{step:MD}
\end{enumerate}
We note the following:
\begin{itemize}
\item In the first step of the first iteration it is necessary to have a starting ``guess'' for $U_e$.
Since $s_a t_{2\beta}$ is assumed to be small, a reasonable starting point is to choose one of the 
eight possibilities 
\begin{equation}
	\tilde{I} = \mbox{diag}(\pm 1, \pm 1, \pm 1)\, .
	\label{eq:Itilde}
\end{equation}
\item In Step~\ref{step:OsOT} we use Eq.~(\ref{eq:OsOT}) to determine $O_n$ and $s_n$. In practice, 
we compute the complex eigenvalues and eigenvectors of the complex, symmetric matrix $S_n$
numerically (and assume that the eigenvalues are non-degenerate).
Then we construct the complex matrix $O_n$ and check that it is approximately orthogonal.
\item Step~\ref{step:Etilde} is the most challenging part of the algorithm.  We have found in practice that 
if an incorrect set of signs has been chosen for $\tilde{I}$, it will
not be possible to solve for $\tilde{E}_n$ (hence the instruction at the end of Step~\ref{step:Etilde}).
\item Once the correct $\tilde{I}$ has been determined,\footnote{Recall that there is an overall
sign ambiguity in $U_e$, so one actually expects two choices for $\tilde{I}$ that yield
solutions for $M_D$.} it is typically sufficient to iterate through
Steps~\ref{step:BnSn}-\ref{step:MD} three to five times in order to determine $U_e$ and $M_D$
to within a reasonable amount of accuracy.  We refer the reader
to Sec.~\ref{sec:results} for further details regarding the accuracy of the method.
\end{itemize}

\subsection{Determination of $\mathbf{\tilde{E}}$}
\label{section:determining-Etilde}

The most challenging step in the algorithm described above is Step~\ref{step:Etilde}, in which
we use Eq.~(\ref{eq:Etildeij}) to solve for the elements of the complex orthogonal matrix $\tilde{E}$.
In this subsection we describe how this may be accomplished.  Throughout this subsection 
we suppress the index $n$ that denotes the iteration number.

As we show in Appendix~\ref{sec:angular-parametrization}, $\tilde{E}$ can typically be parameterized as follows,\footnote{As 
explained in Appendix~\ref{sec:angular-parametrization}, the parameterization in 
Eq.~(\ref{eq:Etilde-euler}) does break down in certain edge
cases (for example, when the 2-2 element of $\tilde{E}$ is equal to unity). 
In most such cases, another angular parameterization could be used.}
\begin{eqnarray}
	\tilde{E} = 
	\left(
		\begin{array}{ccc}
			c_{\eta_1}c_{\eta_3}-c_{\eta_2}s_{\eta_1}s_{\eta_3}\xi & s_{\eta_1}s_{\eta_2} & c_{\eta_1}s_{\eta_3}+c_{\eta_2}c_{\eta_3}s_{\eta_1}\xi \\
			s_{\eta_2}s_{\eta_3}\xi & c_{\eta_2} & -c_{\eta_3}s_{\eta_2}\xi \\
			-c_{\eta_3}s_{\eta_1} - c_{\eta_1}c_{\eta_2}s_{\eta_3}\xi & c_{\eta_1}s_{\eta_2} & c_{\eta_1}c_{\eta_2}c_{\eta_3}\xi -s_{\eta_1}s_{\eta_3} \\
		\end{array}
	\right) \; ,
	\label{eq:Etilde-euler}
\end{eqnarray}
where $c_{\eta_i}\equiv \cos\eta_i$ and $s_{\eta_i}\equiv \sin\eta_i$ ($i=1, 2, 3$), and
where the angles $\eta_1$, $\eta_2$ and $\eta_3$ are assumed to be complex.
The parameter $\xi$ is either $+1$ or $-1$ and is equal to the determinant
of $\tilde{E}$.
The goal in Step~\ref{step:Etilde} is to determine three complex
angles $\eta_i$ and the sign of the discrete parameter $\xi$ such
that Eq.~(\ref{eq:Etildeij}) is satisfied for all $i$ and $j$.
Parameterizing $\tilde{E}$ as in Eq.~(\ref{eq:Etilde-euler}) allows us to solve for the real
and imaginary parts of the $\eta_i$ in a relatively straightforward, prescriptive
manner.

The real and imaginary parts of the $2$-$2$ element of Equation~(\ref{eq:Etildeij}) give the following two relations,
\begin{eqnarray}
	c_{\eta_2}^R & = & \Delta \tilde{B}_{22}^R+c_{\eta_2}^R\cos(2\gamma_2)- c_{\eta_2}^I\sin(2\gamma_2) \label{eq:c2R}\\
	c_{\eta_2}^I & = & \Delta \tilde{B}_{22}^I-c_{\eta_2}^R\sin(2\gamma_2)- c_{\eta_2}^I\cos(2\gamma_2) \label{eq:c2I}\, ,
\end{eqnarray}
in which the $R$ and $I$ superscripts refer to the real and imaginary parts 
[e.g., $c_{\eta_2}^R \equiv \mbox{Re}\!\left(\cos(\eta_2)\right)$].  Equations~(\ref{eq:c2R}) and (\ref{eq:c2I}) are
equivalent to each other, as may be seen by noting that $\sqrt{s} \Delta \tilde{B}$ is anti-Hermitian (see Eq.~(\ref{eq:DeltaB})).
Rewriting Eq.~(\ref{eq:c2R}) in terms of the real and imaginary parts of the complex angle $\eta_2$ yields the following
expression,
\begin{equation}
	\cosh\!\left(\eta_2^I\right)\cos\!\left(\eta_2^R\right)\left[1-\cos(2\gamma_2)\right]
		= \Delta \tilde{B}_{22}^R + \sinh\!\left(\eta_2^I\right)\sin\!\left(\eta_2^R\right) \sin\!\left(2\gamma_2\right) \, ,
\end{equation}
which has the two solutions
\begin{equation}
	\eta_2^R = \alpha - \sin^{-1}\!\left(\frac{\Delta \tilde{B}_{22}^R}{\cal G}\right)
	\label{eq:eta2R1}
\end{equation}
and
\begin{equation}
	\eta_2^R = \alpha - \pi + \sin^{-1}\!\left(\frac{\Delta \tilde{B}_{22}^R}{\cal G}\right)
	\label{eq:eta2R2}
\end{equation}
where
\begin{equation}
	\alpha \equiv \tan^{-1}\!\left(
		\frac{\left(1-\cos(2\gamma_2)\right)\cosh\!\left(\eta_2^I\right)}
		{\sinh\!\left(\eta_2^I\right)\sin\!\left(2\gamma_2\right)}\right)
\end{equation}
and
\begin{equation}
	{\cal G} = \left[
		\left(1-\cos(2\gamma_2)\right)^2\cosh^2\!\left(\eta_2^I\right)
		+ \sinh^2\!\left(\eta_2^I\right)\sin^2\!\left(2\gamma_2\right)
	\right]^{\frac{1}{2}} \,,
\end{equation}
as long as
\begin{equation}
	\sinh^2\!\left(\eta_2^I\right) \geq \frac{\left(\Delta \tilde{B}_{22}^R\right)^2-\left[1-\cos(2\gamma_2)\right]^2}{2\left(1-\cos(2\gamma_2)\right)} \; .
	\label{eq:sinheta2Limits}
\end{equation}

The $1$-$2$ and $2$-$1$ elements of Equation~(\ref{eq:Etildeij}) also give redundant relations, again
due to the fact that $\sqrt{s} \Delta \tilde{B}$ is anti-Hermitian.  The same may be
said for the $2$-$3$ and $3$-$2$ elements.  As a result, we are left
with the following two complex relations,
\begin{eqnarray}
	s_{\eta_1}s_{\eta_2} & = & \Delta \tilde{B}_{12}+s_{\eta_2}^*s_{\eta_3}^*\xi \sqrt{\left|\frac{s_2}{s_1}\right|}e^{-i\left(\gamma_1+\gamma_2\right)} 
		\label{eq:Etilde12}\\
	c_{\eta_1}s_{\eta_2} & = & \Delta \tilde{B}_{32}-s_{\eta_2}^*c_{\eta_3}^*\xi \sqrt{\left|\frac{s_2}{s_3}\right|}e^{-i\left(\gamma_2+\gamma_3\right)} 
		\label{eq:Etilde23}\, .
\end{eqnarray}
The above expressions allow us to express the complex quantities $s_{\eta_1}$ and $c_{\eta_1}$ as follows,
\begin{eqnarray}
	s_{\eta_1} & = & {\cal A} + {\cal B} s_{\eta_3}^* \, ,\label{eq:AB}\\
	c_{\eta_1} & = & {\cal C} + {\cal D} c_{\eta_3}^* \, ,\label{eq:CD}
\end{eqnarray}
where ${\cal A}$, ${\cal B}$, ${\cal C}$ and ${\cal D}$ are functions of $s_{\eta_2}$ and $s_{\eta_2}^*$.
Imposing the constraints $s_{\eta_1}^2 + c_{\eta_1}^2 = s_{\eta_3}^2 + c_{\eta_3}^2 = 1$,
we have
\begin{equation}
	1 = {\cal A}^2 + {\cal C}^2 + {\cal D}^2 +\left({\cal B}^2 - {\cal D}^2\right) \left(s_{\eta_3}^*\right)^2
		+ 2 {\cal A}{\cal B}s_{\eta_3}^* \pm 2 {\cal C}{\cal D} \sqrt{1-\left(s_{\eta_3}^*\right)^2} \, ,
		\label{eq:sineta3star}
\end{equation}
where $c_{\eta_3}^* \equiv \pm \sqrt{1-\left(s_{\eta_3}^*\right)^2}$.  Multiplying
the two expressions in Eq.~(\ref{eq:sineta3star}) by each other leads to a quartic equation
in $s_{\eta_3}^*$, although in practice we typically solve the expressions in Eq.~(\ref{eq:sineta3star})
as stated, since this seems to be more stable numerically.

To summarize, for a given value of $\eta_2^I$ satisfying Eq.~(\ref{eq:sinheta2Limits}) there are
two solutions for $\eta_2^R$ (see Eqs.~(\ref{eq:eta2R1}) and (\ref{eq:eta2R2})).
For both of these we may calculate $\eta_2 = \eta_2^R+ i \eta_2^I$, and so
compute $s_{\eta_2}$ and then ${\cal A}$, ${\cal B}$, ${\cal C}$ and ${\cal D}$.
For each value of $\eta_2$ we may solve Eq.~(\ref{eq:sineta3star}) to obtain a total of 
four solutions for $s_{\eta_3}^*$ and the corresponding values of $c_{\eta_3}^*$.
Back substitution into Eqs.~(\ref{eq:AB}) and (\ref{eq:CD}) then yields $s_{\eta_1}$
and $c_{\eta_1}$.  Thus, for a given value of $\eta_2^I$ satisfying Eq.~(\ref{eq:sinheta2Limits})
we generically expect a total of eight solutions for the sines and cosines of the
complex angles $\eta_1$, $\eta_2$ and $\eta_3$.  It remains to ensure
that the $1$-$1$, $1$-$3$ and $3$-$3$ elements of Eq.~(\ref{eq:Etildeij})
are satisfied.  Recalling the definition for the quantity
$\Delta_{ij}$ in Eq.~(\ref{eq:Deltaij}), we define,
\begin{eqnarray}
	|\Delta|^2 \equiv |\Delta_{11}|^2 + |\Delta_{13}|^2 + |\Delta_{33}|^2 ,
	\label{eq:Deltasq}
\end{eqnarray}
which will generically have eight values for a given value of $\eta_2^I$.  The goal of our algorithm is to 
find value(s) of $\eta_2^I$ (and corresponding values for the various sines and cosines of the complex angles)
that correspond to a zero (or, numerically, a minimum) of Eq.~(\ref{eq:Deltasq}).

Then the method proceeds as follows:
\begin{enumerate}[label=\ref{step:Etilde}(\alph*)]
\item Choose a value for $\xi$ (either $+1$ or $-1$).
\label{step:xi-sign}
\item Choose a value for $\eta_2^I$.
Compute the various combinations of sines and cosines
of the complex angles $\eta_1$, $\eta_2$ and $\eta_3$ that are consistent with the
``central cross'' elements of Eq.~(\ref{eq:Etildeij}) 
[i.e., for $(i,j) = (1,2), (2,1), (2,2), (2,3)$ and $(3,2)$], as described above.
\label{step:complex-angles}
\item Calculate $|\Delta|^2$ for the various combinations of sines and cosines
of the complex angles identified in Step~\ref{step:complex-angles}.\label{step:Delta-squared}
\item Repeat Steps~\ref{step:complex-angles} and \ref{step:Delta-squared}, searching for combinations of the
complex angles that yield $|\Delta|^2\simeq 0$ (in practice, we use an algorithm that searches for a minimum of $|\Delta|^2$).
\item If no solutions are found that satisfy $|\Delta|^2\simeq 0$, return to Step~\ref{step:xi-sign}
and repeat the process for the opposite sign of $\xi$.
\end{enumerate}

\section{Numerical results}
\label{sec:results}

For our numerical analysis we implemented a Monte Carlo algorithm as described in Refs.~\cite{Kiers:2005vx,Kiers:2002cz}
to generate various data sets.
The algorithm scans over random values of the Yukawa matrices defined in Eq.~(\ref{eq:yuk}), searching for
sets of parameters that are consistent with experimental constraints.  
For each set of Yukawa matrices, we were then able to compute $m_e$, $M_\nu$, $M_N$, $M_D$, $U_e$ and $\tilde{E}$. 
We generated, and subsequently analyzed, 16 data sets in this way; 
in each case, the neutrino masses and mixings satisfied the current experimental constraints at the 2$\sigma$ level
as given in Ref.~\cite{Esteban:2018azc}.\footnote{The charged lepton masses generated by the 
routine typically agree with their corresponding experimental values to within a few parts in $10^4$.}  In addition, we verified that the effective neutrino mass for neutrinoless double beta decay is 
always below the present limit~\cite{Dolinski:2019nrj}. For every data set we fixed $\tan\beta = k_2/k_1$ to $3/181$ and 
$v_R = 50$ TeV~\cite{Kiers:2005vx}.\footnote{The authors of Ref.~\cite{Kiers:2005vx} introduce an extra $U$(1) symmetry 
into the left-right model broken by a small dimensionless parameter $\epsilon$. One of the advantages of this 
framework is that it allows scenarios consistent with neutrino phenomenology for a relatively low $v_R$ scale. Since 
we followed this work, we also implemented the $U$(1) horizontal symmetry in our analysis and fixed the value of 
$\epsilon$ to 0.3.  Note that while the horizontal symmetry sets hierarchical scales for the 
Yukawa matrices, it does not impose relations among their elements. Therefore, the inclusion of this symmetry 
does not imply any loss of generality regarding the original problem of unwinding the seesaw.} The remaining 
independent parameters, $s_a$ and $|v_L|$, generally varied within ranges of 
$\cal{O}$($10^{-2} - 1$) and $\cal{O}$($10^{-3} - 10^{-1}$) eV, respectively, whereas the phase of $v_L$ 
took values between 0 and 2$\pi$.  For one of the data sets we purposely set $s_a$ to zero so 
that we could test our method on a parity-conserving data set.
With these choices for the various parameters, the spectrum of the heavy neutrinos spanned
from about 8 to 100 TeV.  Our main focus in this section is not on performing a phenomenological analysis but
on showing that the method described in Sec.~\ref{sec:method-description}
can be successfully applied to data sets that are consistent with experimental constraints, 
allowing us to recover the matrices $U_e$ and $M_D$ in each case. 
It is in this spirit that we are not particularly interested in a phenomenologically inspired 
spectrum for the heavy neutrinos.\footnote{In any case, a lower heavy-neutrino spectrum could in principle
be generated by decreasing the value of $v_R$.}

In the following subsections, we illustrate how the method described in Sec.~\ref{sec:method-description} leads to 
solutions of Eqs.~(\ref{eq26}), (\ref{eq27}) and (\ref{eq31}) for three different scenarios.\footnote{The computer 
code that implements the method is written in \texttt{Mathematica} and is available from the authors
upon request.}  One of the
three respects the generalized parity symmetry in the Dirac Yukawa sector after spontaneous symmetry breaking and the other two
do not.  Of the latter two, one has a relatively
small value of $|s_a|$ (and is thus relatively ``close'' to the parity-conserving limit), while the other
has a larger value of $|s_a|$.  While we only consider these three data sets in detail, we
emphasize that the method was successful 
for all 16 of the data sets.The calculation takes approximately 5 minutes for each dataset on a Desktop PC with Intel Core i7-9700K Processor (8x 3.60 GHz) and 16 GB RAM.

To study these three data sets we show plots of the quantity $|\Delta|^2$, defined in Eq.~(\ref{eq:Deltasq}),
as a function of Im$(\eta_2)$.  Minimizing $|\Delta|^2$
is a key step in determining the elements of the matrix $\tilde{E}$, which then allows us to determine $U_e$ and $M_D$.
For each value of Im$(\eta_2)$ we determine
the sines and cosines of the complex angles in the matrix $\tilde{E}$ (see Eq.~(\ref{eq:Etilde-euler})) that are consistent with the
``central cross'' elements in Eq.~(\ref{eq:Etildeij}).  $|\Delta|^2$ is a measure of how well the 
remaining elements in this equation are satisfied; it is zero for solutions of Eq.~(\ref{eq:Etildeij}).
In principle, minima of $|\Delta|^2$ may need to be found several times, since the 
overall method is iterative.  In the following, we show plots of $|\Delta|^2$ that are obtained
after an appropriate number of iterations have been performed.

\subsection{Parity conserving scenario}

We first consider the parity conserving scenario ($s_a=0$) that was noted above.  For this example,
\begin{equation}
\tilde E = \left(
\begin{array}{ccc}
0 & -1 & 0 \\
-1 & 0 & 0 \\
0 & 0 & 1 \\
\end{array}
\right) \, , ~~~~
U_e = \left(
\begin{array}{ccc}
 -1 & 0 & 0 \\
 0 & 1 & 0 \\
 0 & 0 & 1 \\
\end{array}
\right)
\nonumber
\end{equation}
and
\begin{equation}
\small{
M_D=\left(
\begin{array}{rrr}
1.52\times10^{-5}  & 3.21\times10^{-4} - 3.93\times10^{-5} i   & -6.16\times10^{-4} + 1.97\times10^{-4} i\\ 
-3.21\times10^{-4} - 3.93\times10^{-5} i & 2.62\times10^{-3}   & 1.93\times10^{-3} - 1.67\times10^{-3} i\\ 
6.16\times10^{-4} + 1.97\times10^{-4} i  & 1.93\times10^{-3} + 1.67\times10^{-3} i & 1.81\times10^{-3} \\
\end{array}
\right),
}
\end{equation}
in which we have expressed $M_D$ in units of GeV.
The matrices $\tilde{E}$ and $U_e$ both have a form consistent with one of the expected forms for the parity-conserving case 
(see Section~\ref{sec:derivations},
as well as Ref.~\cite{Senjanovic:2018xtu}).  Also, aside from small numerical errors, $M_D U_e^\dagger$ is Hermitian, as is
expected from Eq.~(\ref{eq26}).
 
\begin{figure}
\begin{center}
\includegraphics[scale= 0.82]{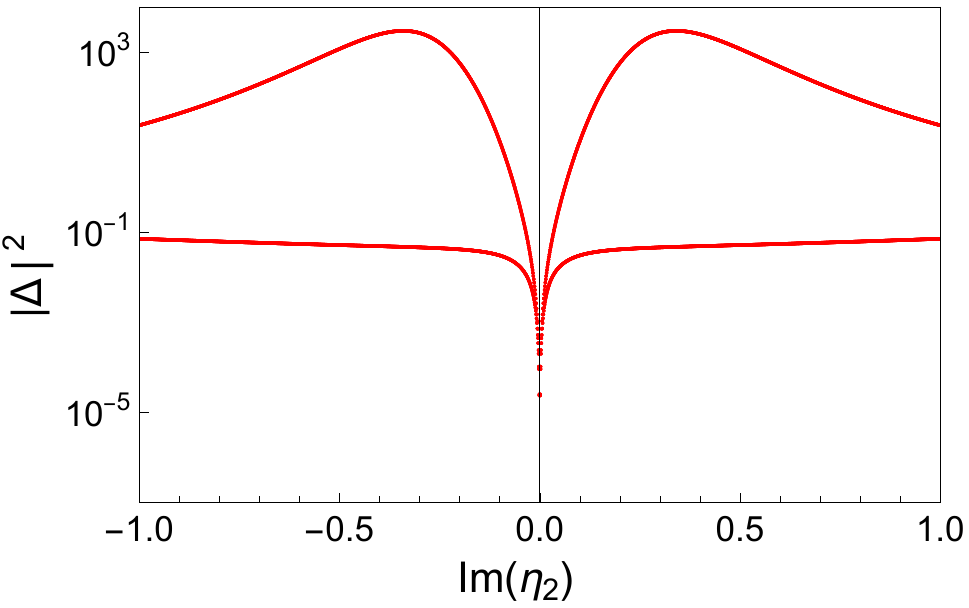}
\end{center}
\caption{$|\Delta|^2$ vs.  Im$(\eta_2)$ for the parity-conserving scenario.}
\label{figure1}
\end{figure}

Figure~\ref{figure1} shows a plot of $|\Delta|^2$ as a function of Im$(\eta_2)$.\footnote{As is 
noted in Sec.~\ref{sec:derivations}, our method
assumes that $\tilde E_{22}\neq\pm1$, a condition that is satisfied for this data set.}
As is evident from the figure, the curves approach zero 
for Im$(\eta_2) \simeq 0$.  We normally expect eight solutions for each value
of Im$(\eta_2)$; to within 
numerical rounding errors, there are two sets of degenerate curves in this case (one set
for each value of Re($\eta_2$) for a given value of Im($\eta_2$)).  
We expect the degeneracies to be removed within parity-violating scenarios. 
In the following subsections we consider two data sets with $s_a \neq 0$.  While both data sets illustrate
the expected breaking of the degeneracy, one of them has
$s_a \ll 1$ and exhibits some qualitative similarities to the parity-conserving scenario.

We note that since this data set has $s_a=0$, we could also use the analytical solution method presented in Section III.A of 
Ref.~\cite{Senjanovic:2018xtu} in this case.  We have analyzed this data set using both the approach of Ref.~\cite{Senjanovic:2018xtu}
and the method described in the present work and have verified that they lead to consistent
values of $U_e$ and $M_D$, up to numerical rounding errors.
Applying the method from Sec.~\ref{sec:method-description} actually required some care in the
parity-conserving scenario.  Technically, in this scenario ${\cal A}$ and ${\cal C}$ 
are both zero in Eq.~(\ref{eq:sineta3star}),
so the equation for $s_{\eta_3}^*$ is quadratic.  In practice we have found that
small numerical errors lead to small but non-zero values for ${\cal A}$ and ${\cal C}$.
Attempting to solve Eq.~(\ref{eq:sineta3star}) as a quartic equation in this case
was not numerically stable, so we resorted to setting ${\cal A}$ and ${\cal C}$
to zero by hand and solving the resulting quadratic equation.

Finally, we note that in the parity-conserving case, our routine also returns values for $M_D$ that
are different than the original one.  When $s_a=0$, it is clear from Eqs.~(\ref{eq26})-(\ref{eq31})
that $-M_D$ is also a solution; this is one of the new solutions that is returned.  Interestingly, however,
another solution emerges that has
\begin{equation}
  \tilde E = \left(
  \begin{array}{ccc}
  0 & 1 & 0 \\
  1 & 0 & 0 \\
  0 & 0 & 1 \\
  \end{array}
  \right) \,  ~~~~
\end{equation}
and, in units of GeV,
\begin{equation}
\small{
M_D=\left(
\begin{array}{rrr}
-2.63\times10^{-4} & -3.74\times10^{-4}-2.64\times10^{-4} i & -2.28\times10^{-4}-1.08\times10^{-4} i\\
3.74\times10^{-4} - 2.64\times10^{-4} i	& -2.24\times 10^{-3} & -1.85\times10^{-3}+ 6.18\times10^{-4} i\\
2.28\times10^{-4}-1.08\times10^{-4} i	& -1.85\times10^{-3}-6.18\times10^{-4} i	& 1.10\times10^{-3} \\
\end{array}
\right),
}
\end{equation}
in which we have dropped small numerical errors.  The negative of the above expression for $M_D$ is also returned.
One can confirm by direct substitution that the new values for $M_D$ are also solutions
of Eqs.~(\ref{eq26}) and (\ref{eq31}).

\subsection{Parity-violating scenario I}

We next consider a scenario with a small degree of parity violation ($s_a=0.00187$).  In this example,
\begin{equation}
\small{
\tilde E = \left(
\begin{array}{rrr}
 -0.0159   -0.0063 i & 1.0000\, +0.0002 i & -0.0134   +0.0212 i \\
  0.9999\, -0.0000 i & 0.0160\, +0.0065 i & -0.0056\, +0.0120 i \\
  0.0052   -0.0117 i & 0.0136\, -0.0214 i & 1.0002\,   +0.0004 i \\
\end{array}
\right) 
}
\label{E2}
\end{equation}
with
\begin{equation}
\small{
U_e=\left(
\begin{array}{rrr}
-1. + 8.09\times10^{-7} i               & -5.20\times10^{-9} - 7.33\times10^{-8} i & 2.22\times10^{-9} + 6.69\times10^{-9} i\\ 
-5.20\times10^{-9} + 7.33\times10^{-8} i & 1. - 5.19\times10^{-7}             i  & 1.41\times10^{-8} + 1.02\times10^{-8} i\\ 
-2.22\times10^{-9} + 6.69\times10^{-9} i & 1.41\times10^{-8} - 1.02\times10^{-8} i & -1. + 5.33\times10^{-7} i\\
\end{array}
\right)
}
\label{eq:Ue-parity-violating-I}
\end{equation}
and, in units of GeV,
\begin{equation}
\small{
M_D=\left(
\begin{array}{rrr}
-4.87\times10^{-6} + 1.58\times10^{-8} i & -1.25\times10^{-4} + 8.83\times10^{-6} i & -1.92\times10^{-4} + 6.38\times10^{-5} i\\ 
 1.25\times10^{-4} + 8.83\times10^{-6} i & -1.79\times10^{-5} + 3.27\times10^{-6} i & -2.76\times10^{-4} + 3.81\times10^{-4} i\\
-1.92\times10^{-4} - 6.38\times10^{-5} i &  2.76\times10^{-4} + 3.81\times10^{-4} i & -1.15\times10^{-3} + 5.50\times10^{-5} i
\end{array}
\right).
}
\label{eq:MD-parity-violating-I}
\end{equation}
Note that the matrix $\tilde E$ in Eq.~(\ref{E2}) has a relatively small value of $\tilde E_{22}$, whereas 
$\tilde E_{12}\approx \tilde E_{21}\approx \tilde E_{33}\approx 1$. All other elements of $\tilde{E}$ are small compared 
to 1. Thus, as one might expect, $\tilde{E}$ is ``close'' to one of the possible parity-conserving forms.
Furthermore, $M_D$ is quite close to being ``sign-Hermitian'' in this example, and $U_e$ is close to 
being a diagonal sign matrix.

\begin{figure}
\begin{center}
\includegraphics[scale= 0.82]{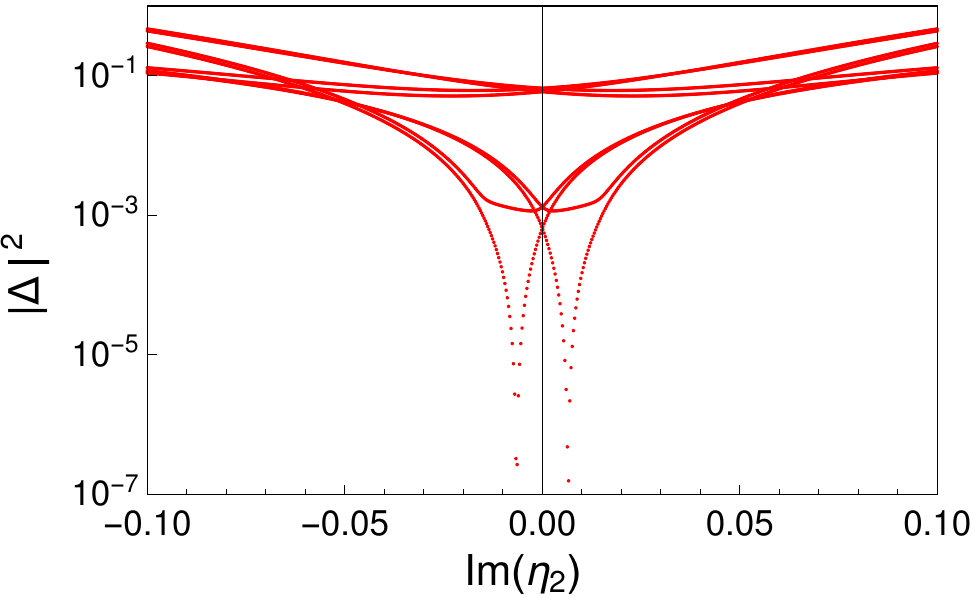}
\end{center}
\caption{$|\Delta|^2$ vs.  Im$(\eta_2)$ for parity-violating scenario I.  Note that the horizontal
and vertical scales are different than those of Figure ~\ref{figure1}.}
\label{figure2}
\end{figure}

We applied our method to this data set and were able to determine the matrices $M_D$ and $U_e$ numerically.
The results are numerically consistent with Eqs.~(\ref{eq:Ue-parity-violating-I}) and (\ref{eq:MD-parity-violating-I}).
Figure~\ref{figure2} shows the corresponding 8-fold family of $|\Delta|^2$ versus Im$(\eta_2)$ curves.
In this semilogarthmic plot, there are slower changing curves that correspond to larger values of $|\Delta|^2$ and 
curves that fall and rise abruptly in the solution region. It is seen that $|\Delta|^2$ approaches zero 
for $|\rm{Im}(\eta_2)|\approx0.0065$.

This case demonstrates several similarities to the one shown above in the parity-conserving scenario. The curves 
still have relatively low $|\Delta|^2$ values and two of them approach $|\Delta|^2 = 0$ at very small values of 
Im$(\eta_2)$.  New features compared to the parity-conserving example are that the 
degeneracies in $|\Delta|^2$ have now been broken and that there are two values of 
$\rm{Im}(\eta_2)$ (located symmetrically about zero) that yield solutions.  These two values of 
$\rm{Im}(\eta_2)$ yield the same solutions for $M_D$, ignoring small numerical errors.
In practice we use the positive root.

\subsection{Parity-violating scenario II}

We now consider a scenario with a larger degree of parity violation ($s_a=-0.51$).  For this example,
\begin{equation}
\small{
\tilde E=\left(
\begin{array}{rrr}
 3.467+2.494 i & 2.623-0.525 i & 1.881-3.866 i \\
 -0.350+0.427 i & 1.052+0.148 i & 0.036-0.162 i \\
 2.540-3.346 i & -0.465-2.628 i & -3.973-1.831 i \\
\end{array}
\right)
}
\end{equation}
with
\begin{equation}
  \small{
    U_e=\left(
\begin{array}{rrr}    
1.-7.82\times 10^{-4} i & 6.25\times 10^{-7}+6.63\times 10^{-5} i & -9.87\times 10^{-6}+5.69\times 10^{-7} i \\
 5.88\times 10^{-7}-6.63\times 10^{-5} i & -1.-2.25\times 10^{-4} i & 2.74\times 10^{-5}+2.56\times 10^{-5} i \\
 -9.87\times 10^{-6}-5.64\times 10^{-7} i & -2.74\times 10^{-5}+2.56\times 10^{-5} i & -1.-1.35\times 10^{-4} i \\
\end{array}
\right)
}
\label{eq:Ue-parity-violating-II}
\end{equation}
and, in units of GeV,
\begin{equation}
\small{
  M_D=\left(
\begin{array}{rrr}  
  5.45\times 10^{-5}-4.32\times 10^{-6} i & 4.12\times 10^{-4}-3.76\times 10^{-6} i & 5.95\times 10^{-5}+1.04\times 10^{-3} i \\
 -4.12\times 10^{-4}-3.78\times 10^{-6} i & -1.31\times 10^{-3}-8.95\times 10^{-4} i & 2.85\times 10^{-3}-3.05\times 10^{-3} i \\
 -5.94\times 10^{-5}+1.04\times 10^{-3} i & 2.85\times 10^{-3}+3.05\times 10^{-3} i & -3.09\times 10^{-3}-1.50\times 10^{-2} i \\
\end{array}
\right).
}
\label{eq:MD-parity-violating-II}
\end{equation}
The $\tilde E$ matrix possesses larger values in this case and loses its resemblance to the $\tilde E$ of the 
parity-conserving case.  The matrices $M_D$ and $U_e$ are still relatively close to being sign-Hermitian and diagonal
sign matrices, respectively, but they are not as close to those forms as were the corresponding expressions
for the ``almost'' parity-conserving case (see Eqs.~(\ref{eq:MD-parity-violating-I}) and 
(\ref{eq:Ue-parity-violating-I}), respectively).

\begin{figure}
\begin{center}
\includegraphics[scale= 0.82]{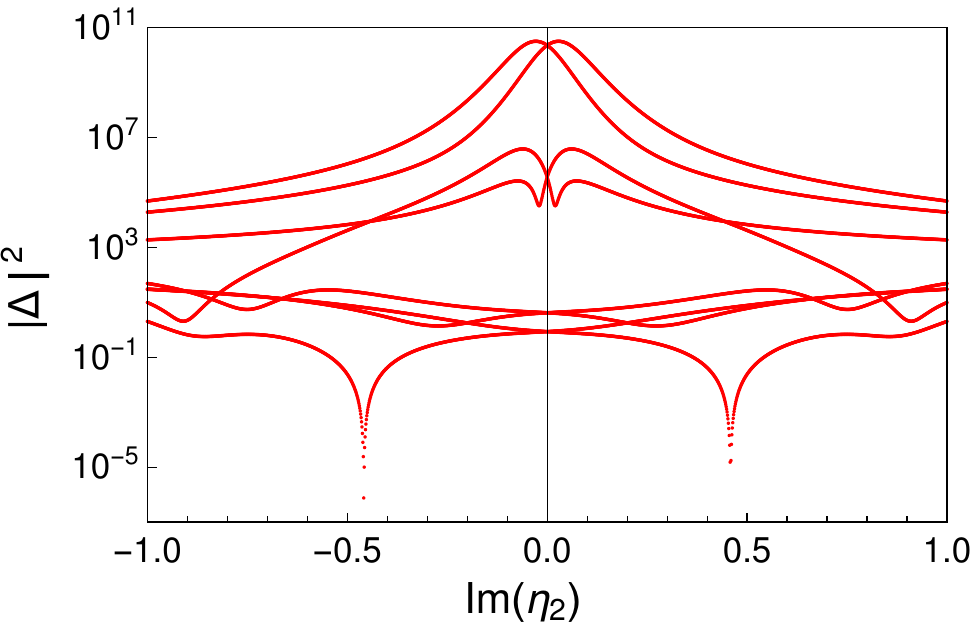}
\end{center}
\caption{$|\Delta|^2$ vs.  Im$(\eta_2)$ for parity-violating scenario II.}
\label{figure3}
\end{figure}

We were able to successfully apply our method to this example and recover values for 
$U_e$ and $M_D$ consistent with Eqs.~(\ref{eq:Ue-parity-violating-II}) and (\ref{eq:MD-parity-violating-II}).
Figure~\ref{figure3} shows the corresponding $|\Delta|^2$ versus Im$(\eta_2)$ curves, with
the solutions evident near $\rm{Im}(\eta_2)\approx\pm0.45$. 
In general, the curves reach significantly larger $|\Delta|^2$ values compared to the previous cases 
and their shapes noticeably differ from the curves of the parity-conserving scenario.

\begin{figure}
\begin{center}
\begin{tabular}{cc}
\includegraphics[scale= 0.82]{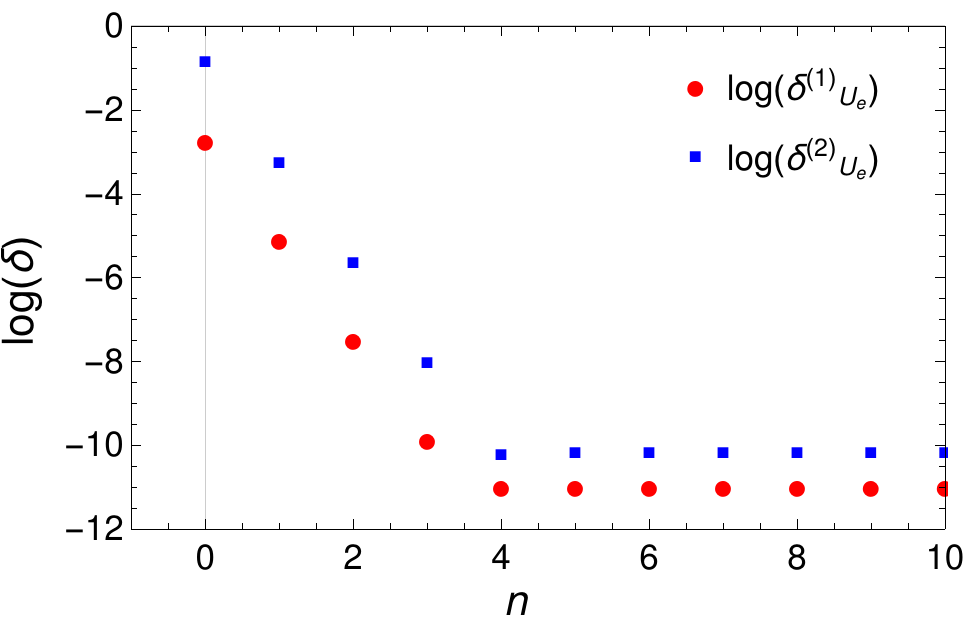}
&
\includegraphics[scale= 0.82]{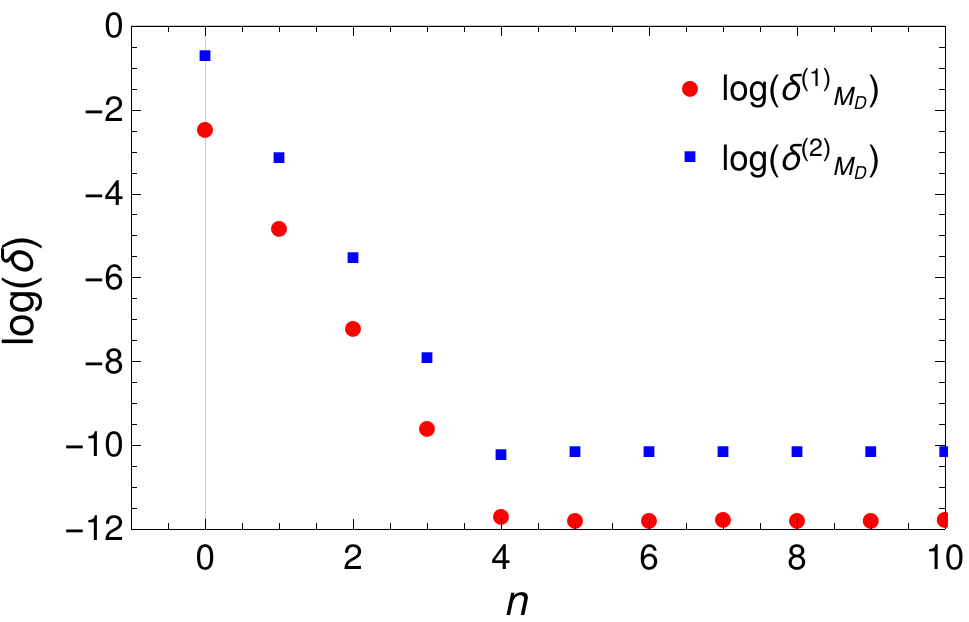}
\end{tabular}
\end{center}
\caption{Convergence plot for $U_e$ (left panel) and $M_D$ (right panel) for parity-violating scenario II.
Please see Eq.~(\ref{eq:delta12}) for the definitions of $\delta^{(1)}_{l}$ and $\delta^{(2)}_{l}$,
with $l=U_e,M_D$.}
\label{figure4}
\end{figure}

We have analyzed the convergence of the method for all of the parity-violating
data sets and the conclusions are similar for all of them. 
As an illustration of that analysis we show the results obtained for the current scenario.
Figure~\ref{figure4} shows two different measures of the relative error between the output of the method for $U_e$ ($M_D$) 
and the true matrix $U_e^0$ ($M_D^0$) as a function of the number of iterations. These measures are defined as
\begin{equation}
\delta^{(1)}_{l} = \mbox{Max} \left( \sqrt{\left(\mbox{Re} \, a^{ij}_l \right)^2 + \left( \mbox{Im} \, a^{ij}_l \right)^2} \right) \, \, \, , \, \, \,
\delta^{(2)}_{l} = \mbox{Max} \left (|\tilde a^{ij}_l|,|\tilde b^{ij}_l| \right)
\label{eq:delta12}
\end{equation}
for $i,j = 1,2,3$ and $l=U_e,M_D$, with
\begin{equation}
\nonumber
a^{ij}_{U_e} =  \frac{\left(U_e-U_e^0 \right)^{ij}}{\left(U_e^0 \right)^{ij}} \, , \, a^{ij}_{M_D} = \frac{\left (M_D-M_D^0\right)^{ij}}{\left (M_D^0\right)^{ij}}  \, ,
\end{equation}
\begin{equation}
\nonumber
\tilde a^{ij}_{U_e} = \frac{\mbox{Re}(U_e-U_e^0)^{ij}}{\mbox{Re}(U_e^0)^{ij}} \, , \, 
\tilde b^{ij}_{U_e} = \frac{\mbox{Im}(U_e-U_e^0)^{ij}}{\mbox{Im}(U_e^0)^{ij}} \, ,
\end{equation}
\begin{equation}
\tilde a^{ij}_{M_D} = \frac{\mbox{Re}(M_D-M_D^0)^{ij}}{\mbox{Re}(M_D^0)^{ij}} \, , \, 
\tilde b^{ij}_{M_D} = \frac{\mbox{Im}(M_D-M_D^0)^{ij}}{\mbox{Im}(M_D^0)^{ij}}
 \, .
\end{equation}
As can be seen from the figure, the convergence is extremely fast, precise and stable for both matrices.

\section{Alternative Methods}
\label{sec:other-methods}

In this section we describe some other numerical methods that we explored in our attempt to solve for $U_e$ and $M_D$.
It is worth noting from the beginning, however, that these alternative methods
were not successful in finding solutions for all 15 of the parity-violating data sets.
Only the method described in Section~\ref{sec:method-description} (or slight variations on that method) 
was successful in this regard.

It is perhaps useful to restate the goal of our analysis.  The equations that define the original
problem are Eqs.~(26), (27) and (31) in Ref.~\cite{Senjanovic:2018xtu}; 
these are stated here as Eqs.~(\ref{eq26}), (\ref{eq27}) and (\ref{eq31}).  Equations~(\ref{eq26}) 
and (\ref{eq27}) come from the relations between $M_D$, $m_e$ and the Yukawa matrices, and Eq.~(\ref{eq31}) 
is the seesaw formula for the light neutrino mass matrix. These equations constitute a system of coupled complex matrix 
equations for which we assume $M_N$ and $M_\nu$ as experimental inputs and $U_e$ and $M_D$ as unknowns to
be solved for.

In the following we briefly describe two
alternative methods that implement different numerical approaches to solve $U_e$ and $M_D$: A) a least squares 
minimization method and B) a fully iterative method.

\subsection{Least squares minimization method}

In our first attempt we employed a least squares minimization technique to solve the three coupled matrix equations 
(Eqs.~(\ref{eq27}) and (\ref{eq31}), together with an equation expressing 
the unitarity condition for $U_e$\footnote{If $s_a t_{2\beta} \neq 0$, Eqs.~(\ref{eq26}) and (\ref{eq27})
are redundant.  Since we are assuming the parity-violating case here,
Eq.~(\ref{eq26}) can be removed from the system.}).  In this method the system 
of $3\times3$ complex matrix equations is transformed into a system of 51 equations for 36 unknown 
matrix elements. The quantity to minimize is the sum of the squared differences between the left-
and right-hand sides of those 51 equations when the solved values for the matrix elements of $M_D$ and 
$U_e$ are substituted. The least squares minimization was achieved employing a variant of the Newton 
method~\cite{newton}.

One complication in this approach is that the equations that we are attempting to solve have
different dimensions -- two of the matrix equations have dimensions of mass, while the third (expressing
the unitarity of $U_e$) is dimensionless.
Moreover, the order of magnitude of the matrix $M_\nu$ is 10$^{-11}$ or 10$^{-12}$ in units 
of GeV compared to 1 in the case of the unitarity condition. In view of this, we employed
three normalization strategies: 1) no dimensional normalization was applied, but 
Eq.~(\ref{eq31}) was multiplied by a dimensionless numerical factor to compensate for the smallness 
of the matrix $M_\nu$; 2) the three equations of the system were transformed so that they have a unit 
matrix on their right-hand side; and 3) the three equations were transformed so that they have 
$m_eM_{\nu}m_e^{-2}$ on their right-hand side. We also considered a hybrid approach in which we 
first applied the third strategy and then used the output for $U_e$ from that approach as a starting
point when using the second strategy.  This approach improved the accuracy in several cases. 
We analyzed these four strategies on the 15 data sets.

The least squares minimization approach to our problem attempts to minimize a particular sum of 
squares while traversing a 36-dimensional space of unknowns.
This approach depends on the initial values chosen for the various unknowns.
In our calculations we set 
the initial values for the matrix $U_e$ to be $U_e=\mbox{diag}(\pm1,\pm1,\pm1)$.
These are reasonable starting points, since the actual $U_e$ matrices for the parity-violating
data sets are still somewhat well approximated by parity-conserving ones with a particular choice of signs.  
We found that this method does not converge when an incorrect 
combination of signs in $\mbox{diag}(\pm1,\pm1,\pm1)$ is used. 
In general, we found that while it
was possible to achieve a solution for all of the data sets using some of the above-mentioned strategies, 
none of these strategies worked for the entire collection of data sets.  
This illustrates the considerable difficulty in solving 
the problem at hand.  By way of contrast,
the method described in Sec.~\ref{sec:method-description} does yield a 
solution for all of the data sets that we studied.

For the cases where the least squares minimization method leads to a solution, we compared the
accuracy achieved to that achieved by the method in Sec.~\ref{sec:method-description},
using $\delta^{(1)}_{U_e}$ as a measure.
The accuracy obtained by the method in Sec.~\ref{sec:method-description} 
turns out to be two orders of magnitude higher on average.
That method also ensures safer control over the solution search since it
reduces the multi-dimensional space of unknown variables to lower dimensional regions at each stage 
of the procedure, whereas the least squares minimization method attempts to find all unknown 
variables at the same time.

\subsection{Fully iterative method}

Given the relatively low accuracy and the instability of the least squares minimization method
described in the previous subsection, 
we have also investigated iterative approaches that are based
on the analytical solution for the parity-conserving case (see Ref.~\cite{Senjanovic:2018xtu}).
In this subsection we describe various attempts along these lines.  
The goal of these approaches is to overcome the limitation
imposed by the fact that $M_D$ is not Hermitian when parity is violated.

We take as our starting point Eqs.~(\ref{eq:OsOT}) and (\ref{eq:Htilde}).  Noting that
$\tilde{H}$ can also be written as
\begin{equation}
\tilde{H}=\frac{1}{\sqrt{M_N}}U_eM_D^\dagger \frac{1}{\sqrt{M_N^*}} \, ,
\end{equation}
we find
\begin{equation}
U_eM_D^\dagger = \sqrt{M_N}O\sqrt{s}\tilde{E}O^\dagger \sqrt{M_N^*} \, ,
\label{UeMDT}
\end{equation}
from which it follows that
\begin{equation}
\tilde{E}=(\sqrt{M_N}O\sqrt{s})^{-1}U_eM_D^\dagger(O^\dagger \sqrt{M_N^*})^{-1} \, .
\label{Eeq}
\end{equation}
We also note that we may write a Riccati equation for $U_e$ that derives from Eq.~(\ref{eq27}),
\begin{equation}
U_e m_e U_e=B_\textrm{\scriptsize Riccati},\,\,\, B_\textrm{\scriptsize Riccati}=is_at_{2\beta}(M_D+e^{-ia}t_\beta m_e)+m_e \;.
\label{riccati}
\end{equation}
A solution of this equation is given by
\begin{equation}
U_e=\sqrt{B_\textrm{\scriptsize Riccati} m_e}m_e^{-1},
\label{rroot}
\end{equation}
where by the notation $\sqrt{B_\textrm{\scriptsize Riccati} m_e}$ we mean the 
principal root of the matrix $B_\textrm{\scriptsize Riccati} m_e$.
Equation~(\ref{riccati}) actually has 8 solutions, which can be constructed by multiplying the principal root
by diagonal matrices $\rm{diag}(\pm 1,\pm 1,\pm 1)$.

The first algorithm that we designed consists of the following steps:
\begin{enumerate}
\item Set initial values for the matrices $U_e$ and $\tilde{E}$ (we take random complex matrices). 
\item Use Eqs.~(\ref{eq:Sdefn}) and (\ref{eq:OsOT}) to calculate the orthogonal matrix $O$ and diagonal matrix $s$.
\item Use Eq.~(\ref{UeMDT}) to calculate $M_D$ using $U_e$ and $\tilde{E}$ set in Step 1, and $O$ and $s$ from Step 2.
\item Calculate $U_e$ by solving Eq.~(\ref{riccati}) (we perform the calculation assuming 8 
possibilities for the roots of the Riccati equation as indicated under Eq.~(\ref{rroot})).
\item Use Eq.~(\ref{Eeq}) to calculate $\tilde{E}$ (after correcting $O$ and $\sqrt{s}$ with the updated $U_e$ 
using Eqs.~(\ref{eq:Sdefn}) and (\ref{eq:OsOT})), where the matrix $U_eM_D^\dagger$ is calculated from Eq.~(\ref{eq26}).
\item Orthogonalize $\tilde{E}$ by iterating $\tilde{E} \to \frac{1}{2}\left(\tilde{E}^T + \tilde{E}^{-1}\right)^T$ repeatedly.
\item Return to Step 2.
\end{enumerate}

We can write this more succinctly using the shorthand notation 
adopted in Section~\ref{sec:method-description}:
\begin{enumerate}
\item Eq.~(\ref{eq31}): $U_{e_i}, \tilde{E}_{_i}  \rightarrow M_{D_i}$ ($i$ stands for the {\it i-th} iteration) 
\item Eq.~(\ref{eq27}): $M_{D_i}  \rightarrow U_{e_{i+1}}$
\item Eq.~(\ref{eq26}): $M_{D_i}, U_{e_{i+1}}  \rightarrow \tilde{E}_{_{i+1}}$
\item Orthogonalize $\tilde{E}_{_{i+1}}$; back to Step 1
\end{enumerate}

Unfortunately, this algorithm ended up being stable for only about the half of the 15 parity-violating data sets
(in some cases we found that even when starting from the solution the algorithm diverges, leading to 
very distant regions in the parameter space). We also tried two different variations, setting 
instead the initial values for $U_e$ and $M_D$ as random complex matrices. These variations can be summarized
as follows,
\begin{enumerate}
\item Eq.~(\ref{eq31}): $M_{D_i}, U_{e_i} \rightarrow \tilde{E}_{_i}$
\item Orthogonalize $\tilde{E}_{_i}$
\item Eq.~(\ref{eq27}): $M_{D_i}  \rightarrow U_{e_{i+1}}$
\item Eq.~(\ref{eq26}): $\tilde{E}_{_i}, U_{e_{i+1}}  \rightarrow M_{D_{i+1}}$; back to Step 1
\end{enumerate}
and 
\begin{enumerate}
\item Eq.~(\ref{eq31}): $M_{D_i}, U_{e_i} \rightarrow \tilde{E}_{_i}$ 
\item Orthogonalize $\tilde{E}_{_{i}}$
\item Eq.~(\ref{eq26}): $\tilde{E}_{_i}, U_{e_{i}}  \rightarrow M_{D_{i+1}}$
\item Eq.~(\ref{eq27}): $M_{D_{i+1}}  \rightarrow U_{e_{i+1}}$; back to Step 1
\end{enumerate}
However, both variations led to results similar to those of the first algorithm, finding solutions for only
about half of the data sets.

We also designed an alternative algorithm that did not use the Riccati equation. 
For this algorithm we only set the initial value for $U_e$ as a random complex matrix.
The algorithm proceeded as follows,
\begin{enumerate}
\item Eq.~(\ref{eq27}): $U_{e_i} \rightarrow M_{D_i}$
\item Eq.~(\ref{eq26}): $M_{D_i}, U_{e_i}  \rightarrow \tilde{E}_{_i}$
\item Orthogonalize $\tilde{E}_{_{i}}$
\item Eq.~(\ref{eq31}): $M_{D_i}, \tilde{E}_{_i}  \rightarrow U_{e_{i+1}}$; back to Step 1
\end{enumerate}
One might expect this algorithm to be more stable, since all four of the steps use the 
{\it i-th} iteration (instead of both the {\it i-th} and ({\it i}+1)-{\it th} iterations at a 
given same step, as in the previous algorithms) and because it depends only on the initial value 
of $U_e$. However, this algorithm did not find a solution for 
any of the data sets.

Finally, we modified one of these algorithms using an approach inspired by the difference-map
algorithm \cite{PhysRevE.78.036706}.  The difference-map algorithm is known to be able to find solutions to iterative mapping
algorithms that are unstable.  Unfortunately, while this approach seemed to show some promise,
our attempts along these lines were also not successful for all of the data sets.

In summary, the fully iterative algorithms were not successful for all
of the data sets. Our experience with these algorithms underscores the
considerable difficulty of solving this system of complex, nonlinear matrix equations
in order to unwind the seesaw mechanism in the parity-violating case.  
Fortunately, the prescriptive method described in Sec.~\ref{sec:method-description}
and illustrated in Sec.~\ref{sec:results} does appear to be able to solve these equations, at least
for all of the data sets considered.

\section{Detailed analysis of the equation $HH^T = S$}

\label{sec:derivations}

In our efforts to extend the ideas of Ref.~\cite{Senjanovic:2019moe} to the parity-nonconserving case, we found it quite helpful to 
first understand the justification for every step in the analytical solution proposed in Ref.~\cite{Senjanovic:2019moe} for the 
parity-conserving case. It became clear to us how essential it is that $H$ be Hermitian in order to derive an 
analytical solution. Yet, in case $H$ is not Hermitian, it nonetheless proved beneficial to factor 
$H = O\sqrt{s}EO^\dagger$ and concentrate on solving for the entries of the complex orthogonal matrix $E$ 
(which will no longer be a simple signed permutation matrix). 

In this section, we thus provide additional context and supporting explanations for the analytical 
derivation carried out in Sec.~III of Ref.~\cite{Senjanovic:2019moe}. Specifically, we will show how to derive the 
matrices in Eq.~(37) of Ref.~\cite{Senjanovic:2019moe} under mild assumptions. We also discuss which of these assumptions 
are necessary and which can be removed. 

The mathematical context is the following: we assume $H$ is an unknown, Hermitian $3\times 3$ matrix 
and that $S = HH^T$ is known. The goal is to solve for $H$ given $S$. As in Eq.~(27) of Ref.~\cite{Senjanovic:2019moe}, 
we assume that $S = OsO^T$ has been placed in ``symmetric normal form''; here $O$ is a complex orthogonal 
matrix and $s$ is block diagonal with symmetric Jordan blocks. Precise details can be found 
in Section~XI.3 of Ref.~\cite{Gantmacher1960}; we will only consider the case when $s$ is diagonal for 
the sake of simplicity. 

The central claim is that $H=O\sqrt{s}EO^\dagger$, where $E$ is a signed permutation matrix 
whose form is determined by $s$. In fact, even in the parity-violating case (when $H$ is not Hermitian), 
one can decompose $H$ as $H=O\sqrt{s}EO^\dagger$ for some complex orthogonal matrix $E$, as the next 
lemma shows. Any further specification of $E$ is precisely linked to the assumption that $H$ is Hermitian.  

\begin{lemma}\label{siphon}
Suppose that $H$ is an invertible complex square matrix (not necessarily Hermitian), $O$ is a 
complex orthogonal matrix, and $s$ is a diagonal matrix such that $HH^T = OsO^T$. Then for any 
choice of square root $\sqrt{s}$, there exists a complex orthogonal matrix $E$ such that $H = O\sqrt{s}EO^\dagger$. 
\end{lemma}

We begin by observing that 
$$
HH^T = O\sqrt{s} (O\sqrt{s})^T;
$$
therefore 
$$
I = (H^{-1}O\sqrt{s}) (H^{-1}O\sqrt{s})^T,
$$
so $H^{-1}O\sqrt{s}$ is some complex orthogonal matrix, say $P^{-1}$. Rearranging, we have 
\begin{align*}
H^{-1}O\sqrt{s} &= P^{-1}\\
O\sqrt{s}P &= H.
\end{align*}

Note that $O^\dagger$ is a complex orthogonal matrix since $O$ is. Thus if we define 
$E = P(O^{-1})^\dagger$, $E$ is once again complex orthogonal and $P=EO^\dagger$. 
Thus $H = O\sqrt{s}EO^\dagger$ for a complex orthogonal matrix $E$. 

In addition to the assumption that $s$ is diagonal (which is made in Section III.D(i) of Ref.~\cite{Senjanovic:2019moe}), 
we further assume that
\begin{enumerate}
\item[(1)] $s$ has nonzero eigenvalues,
\item[(2)] $s$ has distinct eigenvalues.
\end{enumerate}
All of the above assumptions are mild in a probabilistic sense: they hold with probability $1$ in the 
space of all possible matrices $H$. 
Later we will show that assumption (2) is necessary for the analytical solution described in Ref.~\cite{Senjanovic:2019moe}, 
for otherwise one can exhibit infinitely many matrices $H$ with $HH^T = S$.  In contrast, we show that 
assumption (1) is not necessary. 

\subsection{Determination of $E_I$, $E_{II}$} 
The number of nonreal eigenvalues of $s$ is even; moreover such eigenvalues come in complex conjugate 
pairs. As mentioned in Eq.~(24) and following of Ref.~\cite{Senjanovic:2019moe}, this is due to the characteristic polynomial of 
$S = HH^T$ having real coefficients whenever $H$ is a Hermitian matrix. 

Under the assumption that $s$ is of size $3\times 3$, there are thus only two cases: 
\begin{enumerate}
\item[(I)] All eigenvalues are real: $s = \operatorname{diag}(s_0,s_1,s_2)$, each $s_i\in \mathbb{R}$. 
\item[(II)] There is one pair of complex conjugate eigenvalues: $s = \operatorname{diag}(z,s_0,z^*)$, $s_0\in \mathbb{R}$. 
\end{enumerate}
In either case, one can find a diagonal matrix $\sqrt{s}$ such that $\sqrt{s}\sqrt{s} = s$. Moreover, 
one may assume that the complex entries of $\sqrt{s}$ come in conjugate pairs. 

\begin{proposition}\label{p1}
Assume that $s$ is diagonal, satisfies (1) and (2), and has eigenvalues ordered as in (I) or (II) above. 
Then $H$ must be equal to one of the matrices 
$$
O\sqrt{s}E O^\dagger,
$$
where $E$ comes from the finite list of possibilities: 
$$
E = \left(
\begin{array}{ccc}
\pm1 & 0 & 0 \\ 
0 & \pm1 & 0 \\
0 & 0 & \pm1
\end{array}
\right)
$$
in case (I) and 
$$
E = \left(
\begin{array}{ccc}
0 & 0 & \epsilon \\ 
0 & \pm1 & 0 \\
\epsilon & 0 & 0
\end{array}
\right)
$$
in case (II), here $\epsilon = \pm1$ and we are just emphasizing that the two corner entries must be equal. 
\end{proposition}

By Lemma \ref{siphon}, we can write $H = O \sqrt{s} EO^\dagger$ for some complex orthogonal matrix $E$. 
Now we observe the following: 
\begin{align*}
H&=O\sqrt{s}EO^\dagger \\
H^\dagger&=O(\sqrt{s}E)^\dagger O^\dagger;
\end{align*}
therefore $H$ is Hermitian if and only if $\sqrt{s} E$ is. Since we assumed $H$ was Hermitian, we get the following equations: 
\begin{align*}
\sqrt{s}E &= E^\dagger \sqrt{s}^*\\
\sqrt{s}E {\sqrt{s}^*}^{-1} &= E^\dagger
\end{align*}

The last equation can be rewritten as 
\begin{align}\label{eq1}
\sqrt{s}E {\sqrt{s}^*}^{-1} =  {E^*}^{-1}
\end{align}
since $E^T = E^{-1}$. Conjugating Eq.~(\ref{eq1}) yields
$$
\sqrt{s}^*  E^* \sqrt{s}^{-1} = E^{-1},
$$
and now taking the inverse of both sides we get 
\begin{align}\label{eq2}
\sqrt{s} {E^*}^{-1} {\sqrt{s}^*}^{-1} = E.
\end{align}

Combining Eqs.~(\ref{eq1}) and (\ref{eq2}), we have 
$$
sE {s^*}^{-1} = \sqrt{s} (\sqrt{s}E {\sqrt{s}^*}^{-1}){\sqrt{s}^*}^{-1} = \sqrt{s} {E^*}^{-1} {\sqrt{s}^*}^{-1} = E.
$$
Therefore $E$ commutes with $s$, but up to a conjugation. But this can be remedied, for we know that $s^*$ is
a diagonal matrix whose entries are just a permutation of the diagonal entries of $s$. In other words, there 
exists a permutation matrix $Q$ (possibly the identity matrix) such that 
$
s^* = Q s Q^{-1}
$, which implies ${s^*}^{-1} = Q s^{-1} Q^{-1}$ by taking the inverse of both sides. 
As with all permutation matrices, $Q$ is (real) orthogonal: $Q^T = Q^{-1}$. Putting this all together, we have 
\begin{align*}
s EQ s^{-1} Q^{-1} &= E, \\
s (EQ) s^{-1} &= EQ,
\end{align*}
so the complex orthogonal matrix $EQ$ commutes with $s$. 

Since $s$ has distinct eigenvalues, $EQ$ must be diagonal. The only diagonal $3\times 3$ complex orthogonal 
matrices $A$ are the eight possibilities for 
$$
A = \left(
\begin{array}{ccc}
\pm1 & 0 & 0 \\ 
0 & \pm1 & 0 \\ 
0 & 0 & \pm 1
\end{array}
\right),
$$
because of the requirement that $A^2 = A A^T = I$. 

Finally, to establish the form of the matrix $E$, and thus find all solutions $H$, we just analyze what 
$Q$ was in cases (I) and (II), respectively. 

\begin{enumerate}
\item[(I)] If $s = \operatorname{diag}(s_0,s_1,s_2)$ with each $s_i\in \mathbb{R}$, then $s = s^*$ already, 
so $Q = I$. Therefore $E$ itself must be one of the $8$ possibilities 
$$
E =  \left(
\begin{array}{ccc}
\pm1 & 0 & 0 \\ 
0 & \pm1 & 0 \\ 
0 & 0 & \pm 1
\end{array}
\right).
$$
Moreover, all of these possibilities are realizable -- that is, $O\sqrt{s}EO^\dagger$ is Hermitian in each 
case\footnote{Note that this relies on the entries of $s$ being nonnegative. Indeed, one can show 
(through a somewhat technical case-by-case argument) that if $s$ has at least one negative eigenvalue, 
then this eigenvalue is repeated, violating assumption (2).}. 

\item[(II)] If $s = \operatorname{diag}(z,s_0,z^*)$, then 
$$
Q = \left(
\begin{array}{ccc}
0 & 0 & 1 \\ 
0 & 1 & 0 \\ 
1 & 0 & 0
\end{array}
\right),
$$
so 
$$
E =  \left(
\begin{array}{ccc}
0 & 0 & \pm 1 \\ 
0 & \pm1 & 0 \\ 
\pm1 & 0 & 0 
\end{array}
\right).
$$
However, only half of these are valid possibilities yielding $O\sqrt{s}EO^\dagger$ Hermitian. 
Indeed, let us write 
$$
E =  \left(
\begin{array}{ccc}
0 & 0 & \epsilon_1 \\ 
0 & \epsilon_2 & 0 \\ 
\epsilon_3 & 0 & 0 
\end{array}
\right),
$$
where each $\epsilon_i = \pm1$. Then 
$$
\sqrt{s}E = \
\left(
\begin{array}{ccc}
0 & 0 & \epsilon_1 \sqrt{z} \\ 
0 & \epsilon_2 \sqrt{s_0} & 0 \\ 
\epsilon_3 \sqrt{z}^*& 0 & 0 
\end{array}
\right),
$$
which is Hermitian if and only if $\epsilon_1 = \epsilon_3$. So $E$ must take on the form 
$$
E =  \left(
\begin{array}{ccc}
0 & 0 & \epsilon \\ 
0 & \pm1 & 0 \\ 
\epsilon & 0 & 0 
\end{array}
\right),
$$
where $\epsilon = \pm 1$. 
\end{enumerate}

\subsection{Distinct eigenvalues are necessary}

We want to point out that the assumption (2) of Proposition \ref{p1} concerning distinct 
eigenvalues is necessary in order to find only finitely many solutions for $H$. 

\begin{proposition}
If $s$ is a diagonal matrix of dimension at least $2$ with a repeated eigenvalue, then the equation 
$HH^T = OsO^T$ has infinitely many solutions (given $O$,$s$) with Hermitian $H$. 
\end{proposition}

It suffices to consider only the equation $H_0H_0^T = s$ by performing the change of variables 
$H_0 = O^{-1}H(O^{-1})^\dagger$. Indeed, $H_0$ is Hermitian if and only if $H$ is. Moreover, seeing as
\begin{align*}
H_0H_0^T & = O^{-1} H (O^{-1})^\dagger (O^{-1})^* H^T (O^{-1})^T\\
&= O^{-1} H H^T (O^{-1})^T,
\end{align*}
we find that $H_0H_0^T = s$ if and only if $H H^T = OsO^T$. 

First suppose $s$ has a repeated real eigenvalue $s_0$; up to permutations assume 
$$s = \operatorname{diag}(s_0,s_0,s_1,\hdots)$$ (the later eigenvalues may be real or come in 
complex conjugate pairs). We already know we can find a Hermitian matrix $H_1$ such that 
$H_1H_1^T = \operatorname{diag}(s_1,\hdots)$ (so $H_1$ has dimension $2$ smaller). Therefore 
it suffices to show that there are infinitely many Hermitian $2\times 2$ matrices $A$ such that 
$$
\left(
\begin{array}{c|c}
A & 0 \\\hline
0 & H_1
\end{array}
\right)
\left(
\begin{array}{c|c}
A & 0 \\\hline
0 & H_1
\end{array}
\right)^T = s;
$$
i.e., such that $AA^T = \left(\begin{array}{cc} s_0 & 0 \\ 0 & s_0 \end{array}\right)$. 

Set 
$$
A = \left(\begin{array}{cc}
a & i b \\ 
-ib & a 
\end{array}\right),
$$ with both $a,b\in \mathbb{R}$. Then as long as $a^2-b^2 = s_0$, $A$ is such a solution. The 
equation $a^2-b^2 = s_0$ has infinitely many solutions for $a,b$ (graphically, the points on 
that hyperbola); for example, if $s_0$ is nonnegative, then $a = \pm \sqrt{s_0+b^2}$ for any 
choice of $b$ gives a distinct solution. 

Second, suppose $s$ has a repeated complex (not real) eigenvalue $z$. Up to permutations assume 
$$
s = \operatorname{diag}(z,z, z^*,z^*,s_0,\hdots).
$$
Once again it suffices to find infinitely many Hermitian matrices $A$ such that $AA^T = \operatorname{diag}(z,z, z^*,z^*)$. 

Set 
$$
A = 
\left(
\begin{array}{cccc}
0 & 0 & x & y \\ 
0 & 0 & -y & x \\
\bar x & -\bar y & 0 & 0 \\
\bar y & \bar x & 0 & 0
\end{array}
\right).
$$
A straightforward calculation reveals that, as long as $x^2+y^2 = z$, $AA^T = \operatorname{diag}(z,z, z^*,z^*)$. 
Again there are infinitely many solutions to the equation $x^2+y^2 = z$ (pick any $x\in \mathbb{C}$ and 
there is at least one solution for $y$). 

\subsection{Handling an unrepeated eigenvalue $0$}

Suppose $HH^T = OsO^T$ as before with $s$ diagonal and having distinct eigenvalues (assumption (2)). 
Now suppose that $s$ has $0$ as one of its eigenvalues. Without loss of generality, we assume that 
$HH^T = s$ (using a change-of-basis as before) and that $s = \operatorname{diag}(a,b,0)$, where $a\ne 0, b\ne 0$. 

\begin{proposition}
Under the above assumptions, $H$ must be of the form 
$$
\left(
\begin{array}{c|c}
\bar H & 0 \\\hline
0 & 0
\end{array}
\right).
$$
\end{proposition}

Let $e_3$ denote the standard column vector $\left(\begin{array}{c} 0 \\ 0 \\ 1 \end{array}\right)$. 

Note that the nullspace of $s^*$ is spanned by $e_3$. Set $v = H^Te_3$. Then $Hv = 0$ since $se_3 = 0$, 
so $H^THv = 0$ as well. Recalling that $s^* = H^*H = H^TH$, we see that $s^*v = 0$. Thus $v = ce_3$ for 
some scalar $c$. Now observe that 
\begin{align}\nonumber
H^*e_3 &= ce_3\\\label{c=0}
He_3 &= c^*e_3.
\end{align}
Since $0 = H(ce_3) = |c|^2e_3$, we must have $c = 0$. Eq. (\ref{c=0}) together with $c=0$ implies that $H$ is a Hermitian matrix whose last column 
is all $0$'s. In other words, 
$$
H = \left(
\begin{array}{c|c}
\bar H & 0 \\\hline
0 & 0
\end{array}
\right)
$$
for some Hermitian $2\times2$ matrix $\bar H$. 

Therefore the problem of solving for $H$ reduces to solving for $\bar H$, and the smaller system
$$
\bar H \bar H^T = \operatorname{diag}(a,b)
$$ can be solved as before.

\section{Discussion and Conclusions}
\label{sec:discussion-conclusions}

It is well known that at least two neutrinos are light massive particles. However, it remains 
crucial to understand the origin of neutrino mass. The seesaw mechanism is a compelling
possibility but it needs to be probed as a consistent explanation.
For the charged fermions, experiments support the SM explanation that ties
the particles' masses to the corresponding Yukawa terms in the underlying
theory.  In the case of neutrinos, one would want to be able to determine 
the Dirac mass ($M_D$) between the left-handed neutrinos and 
the new neutral lepton singlets ($N$) in terms of the light neutrino masses and mixings ($M_\nu$) 
and the mass matrix of the heavy states ($M_N$). Therefore, probing the seesaw requires the 
measurement of these two matrices and a scheme that allows the determination of $M_D$ from them.

Within the context of minimal extensions of the SM, $M_D$ 
is not unambiguously determined in terms of $M_\nu$ and $M_N$.  This problem can be overcome,
however, by including more structure in the theory.  This is precisely the case for the left-right 
symmetric model, where the seesaw is a natural outcome of spontaneous symmetry breaking. 
In this paper we have studied the scenario in which the left-right symmetry is implemented
via a discrete generalized parity, $\cal{P}$.  In this scenario one can solve
for $M_D$ analytically using the approach described in Refs.~\cite{Senjanovic:2018xtu,Senjanovic:2019moe}
as long as the bidoublet Higgs field has a real vacuum expectation value.  In contrast, 
for a complex VEV, which induces $\cal{P}$ parity violation in the Dirac Yukawa sector, the problem 
is more difficult to handle and an analytical solution is lacking. Although this case can in 
principle be addressed numerically, we are not aware of any defined numerical procedure in the 
literature. With the intention of filling this gap, the goal of this paper has been to design and 
test a prescriptive numerical method that allows one to determine $M_D$ only from the physical information 
contained in the matrices $M_\nu$ and $M_N$.

For the parity-violating case, we have found that the problem of determining $M_D$ 
is inherently tied to the knowledge of $U_e$, the unitary matrix associated with the transformation 
from the gauge basis to the charged-diagonal basis. Therefore, both matrices need to be solved 
simultaneously. The method proposed in this paper, as described in Sec.~\ref{sec:method-description}, fulfills 
this goal through an iterative procedure that has proven to be stable and has led to solutions for all of the 
tested data sets.  We illustrated the procedure explicitly in Sec.~\ref{sec:results} for three different
data sets that had varying degrees of parity violation.

Finally, it is worth stressing the difficulty of the problem faced in this article. As a matter of fact, 
in Sec.~\ref{sec:other-methods} we presented alternative iterative methods that were stable for some 
data sets while not for others. The problem of probing the seesaw mechanism when parity is violated 
is important and challenging enough to require a robust solution method.

\section*{Acknowledgements}

The authors wish to thank G. Senjanovi\'{c} for helpful communication and M. Assis and D. Simons 
for permission to use their computer code.  The authors also thank T. Lehrian and W. Slauson for technical assistance. K.K. thanks Taylor University for financial support
and for support during his sabbatical. T.T. and A.S. thank CONICET and ANPCyT (under projects PICT 2017-2751 and
PICT 2018-03682).

\appendix

\section{Connection to Notation in Ref.~\cite{Kiers:2005vx}}

\label{app:model}

In this
appendix we outline the procedure for diagonalizing the charged and
neutral mass matrices and we also specify the relations between
the charged-diagonal basis (used in Ref.~\cite{Senjanovic:2018xtu} and in the present work)
and the gauge basis (used in Ref.~\cite{Kiers:2005vx} and summarized at the beginning of Sec.~\ref{sec:model}).

Equations (\ref{eq:mlep}), (\ref{eq:MLR}), (\ref{eq:MLL}), (\ref{eq:MRR})
and (\ref{eq:Mnu-2005})
give the explicit expressions for $M_\ell$, $M_{LR}$, $M_{LL}$, $M_{RR}$
and the light neutrino mass matrix, respectively,
in terms of Yukawa coupling matrices in the gauge basis.  
The charged lepton mass matrix may be diagonalized using
a biunitary transformation as follows,
\begin{eqnarray}
	m_e \equiv M_\ell^\textrm{\scriptsize diag} = V_L^{\ell\dagger}
		M_\ell V_R^\ell \; ,
		\label{eq:diagchargedlep}
\end{eqnarray}
where the elements of $m_e$ are taken to be real
and positive.  The light and heavy neutrino mass matrices 
may also be diagonalized using unitary matrices,
\begin{eqnarray}
	M_\nu^\textrm{\scriptsize diag} &=& 
		V_L^{\nu \dagger} \left(M_{LL}^\dagger-M_{LR}M_{RR}^{-1}M_{LR}^T \right) V_L^{\nu *}  \; ,
	\label{eq:mnudiag}\\
	M_R^\textrm{\scriptsize diag} &=&
		V_R^{\nu T}M_{RR}V_R^\nu \; .
	\label{eq:mrdiag}
\end{eqnarray}
The unitary matrices used to diagonalize the charged and neutral lepton mass matrices
are then used to construct the left- and right-handed PMNS matrices,
\begin{eqnarray}
	V_L &=& B_\phi^\dagger V_L^{\ell\dagger}V_L^\nu S_L \; ,
		\label{eq:MNStilde1}\\
	V_R &=& B_\phi^\dagger V_R^{\ell\dagger}V_R^\nu S_R \; ,
		\label{eq:MNStilde2}
\end{eqnarray}
where $B_\phi$ is a diagonal phase matrix and $S_L$ and $S_R$ are diagonal
sign matrices; these diagonal matrices are used to bring 
$V_L$ and $V_R$ into
their conventional forms.  Defining $\nu_{L,R}$
and $e_{L,R}$ to be the neutral and charged lepton fields in the mass
basis (i.e., in the basis in which the mass matrices are diagonal), we have
\begin{eqnarray}
	\nu_{L,R} &=& S_{L,R}^\dagger V_{L,R}^{\nu\dagger}\nu_{L,R}^\prime
		\; , \\
	e_{L,R} &=& B_\phi^\dagger V_{L,R}^{\ell\dagger}e_{L,R}^\prime \; ,
\end{eqnarray}
where $\nu_{L,R}^\prime$ and $e_{L,R}^\prime$ are the corresponding
fields in the gauge basis.
The left- and right-handed PMNS matrices
appear in the charged-current Lagrangian when it is written in terms
of the fields in the mass basis,
\begin{eqnarray}
	{\cal L}_{CC} \simeq -\frac{g}{\sqrt{2}}\overline{e}_L V_L
		\gamma_\mu\nu_L W_L^{\mu -}
		-\frac{g}{\sqrt{2}}\overline{e}_R V_R
		\gamma_\mu\nu_R W_R^{\mu -}+ \textrm{h.c.}
\end{eqnarray}

Finally, we note that the left-handed PMNS matrix is parameterized as follows 
in Ref.~\cite{Kayser:2002qs} (and in Ref.~\cite{Kiers:2005vx}),
\begin{eqnarray}
	V_L = {\cal U}^{(0)}(\theta_{12},\theta_{23},
		\theta_{13},\delta_L) A_L,
		\label{eq:ULMNS}
\end{eqnarray}
where $A_L$ is a diagonal matrix that may be 
written in terms of two Majorana phases, 
$A_L=\textrm{diag}(e^{i\alpha_1/2},e^{i\alpha_2/2},1)$.
The matrix ${\cal U}^{(0)}$ may be written as
\begin{eqnarray}
	{\cal U}^{(0)}(\theta_{12},\theta_{23},\theta_{13},\delta_L) & & 
		\nonumber \\
		& &\!\!\!\!\!\!\!\!\!\!\!\!\!\!\!\! =  \left(\begin{array}{ccc}
		c_{12}c_{13} & s_{12}c_{13} & s_{13}e^{-i\delta_L} \\
		-s_{12}c_{23}-c_{12}s_{23}s_{13}e^{i\delta_L} &
			c_{12}c_{23}-s_{12}s_{23}s_{13}e^{i\delta_L} &
			s_{23}c_{13} \\
		s_{12}s_{23}-c_{12}c_{23}s_{13}e^{i\delta_L} &
			-c_{12}s_{23}-s_{12}c_{23}s_{13}e^{i\delta_L} &
			c_{23}c_{13} \\
		\end{array}\right) \! ,
	\label{eq:U_zero_mns}
\end{eqnarray}
where $s_{ij}$ and $c_{ij}$ refer to the sines and cosines, respectively,
of the (real) angles $\theta_{12}$, $\theta_{13}$ and 
$\theta_{23}$.
The interested reader is referred to Ref.~\cite{Kiers:2005vx}
for a parameterization of $V_R$.

Reference~\cite{Senjanovic:2018xtu} works in a basis in which the 
charged lepton mass matrix is diagonal, but the neutrino mass
matrices are not; we have called this basis the ``charged-diagonal'' basis.
The diagonalization of the charged lepton mass matrix
was shown above, in Eq.~(\ref{eq:diagchargedlep}).
Equations~(\ref{eq26})-(\ref{eq31}) also include the matrix $U_e$;
this matrix is defined in terms of the matrices $V_L^{\ell}$ and $V_R^\ell$
that are used to diagonalize $M_\ell$,
\begin{eqnarray}
	U_e = B_\phi^\dagger V_R^{\ell\dagger} V_L^{\ell} B_\phi \, .
	\label{eq:Ue-def}
\end{eqnarray}

It is straightforward to work out the 
relations between $M_\nu$, $M_N$ and $M_D$ (which are defined in the charged-diagonal basis)
and $M_{LL}^\dagger-M_{LR}M_{RR}^{-1}M_{LR}^T$,
$M_{RR}$ and $M_{LR}$ (which are defined in the gauge basis).
The specific relations are as follows,
\begin{eqnarray}
	M_\nu & = & B_\phi \left(V_L^{\ell}\right)^T \left(M_{LL}^\dagger-M_{LR}M_{RR}^{-1}M_{LR}^T\right)^*V_L^{\ell} B_\phi \;, 
	\label{eq:Mnu_relation}\\
	M_N & = & B_\phi^\dagger V_R^{\ell \dagger} M_{RR}^*V_R^{\ell *} B_\phi^\dagger \; ,\\
	M_D & = & B_\phi^\dagger V_R^{\ell \dagger} M_{LR}^\dagger V_L^{\ell} B_\phi \;.
\end{eqnarray}

\section{Numerical Determination of $U_e$ for a Given ${\cal M}$}

\label{app:Ue-determination}

In this appendix we describe a procedure that can be used
to determine $U_e$ with a high degree of accuracy 
once an approximate expression for
${\cal M}$ has been determined in Step~\ref{step:calM}
of the procedure described in Sec.~\ref{subsec:iterative-method};
the procedure described here is used in Step~\ref{step:Ue} in that section.
Our approach assumes that $s_a t_{2\beta}$ is somewhat
small, in which case $U_e$ is close to diagonal.\footnote{
The authors of Ref.~\cite{Senjanovic:2018xtu} outlined an alternative approach for
estimating $U_e$, which is to express it in terms of a
series expansion in powers of $s_a t_{2\beta}$.}
One potential point of confusion is that the method described in this Appendix is an
iterative one that is itself used in the context of another iterative procedure
(i.e., the one described in Sec.~\ref{subsec:iterative-method}).  In this
Appendix we will suppress the index for the ``Step \#'' in the larger
iterative process.  The index $m$ that is used here refers to the
$m$th step in the iterative process used to determine $U_e$ for a given (i.e., fixed) step
of the procedure described in Sec.~\ref{subsec:iterative-method}.

We start by recalling the definition of ${\cal M}$ in Eq.~(\ref{eq:calM}), which allows
us to reexpress Eq.~(\ref{eq27}) as follows,
\begin{eqnarray}
    U_e m_e - m_e U_e^\dagger = i s_a t_{2\beta}{\cal M} \, ,
    \label{eq:Ueme}
\end{eqnarray}
where $m_e$ is a diagonal matrix containing the charged lepton masses.
In the limit that $s_a t_{2\beta}$ goes to zero, the unitary matrix $U_e$ becomes a diagonal matrix
whose non-zero entries are $\pm 1$.
With this in mind, we define the matrix $U_e^{(m)}$ in terms of a product of $m+1$ unitary 
matrices $U^{(j)}$,
\begin{eqnarray}
    U_e^{(m)} = \prod_{j=0}^{m} U^{(j)} = U^{(m)} U^{(m-1)} \cdots U^{(1)} U^{(0)} \;,
    \label{eq:Ue-product-truncated}
\end{eqnarray}
where $U^{(0)} = \tilde{I}$ is the diagonal sign matrix defined in Eq.~(\ref{eq:Itilde}).
The matrix $U_e$ is taken to be the limit of Eq.~(\ref{eq:Ue-product-truncated})
as $m$ approaches infinity,
\begin{eqnarray}
    U_e = \lim_{m\to\infty} U_e^{(m)} = \prod_{j=0}^{\infty} U^{(j)} \; .
    \label{eq:Ue-product}
\end{eqnarray}
Each of the unitary matrices $U^{(j)}$ may be expressed as follows,
\begin{equation}
	U^{(j)} = \exp\!\left(\sum_{i=1}^9\frac{i \alpha_i^{(j)}}{2} \lambda_i \right) 
        = 1 + \left(\sum_{i=1}^9\frac{i \alpha_i^{(j)}}{2} \lambda_i \right)  
        + \frac{1}{2!} \left(\sum_{i=1}^9\frac{i \alpha_i^{(j)}}{2} \lambda_i \right)^2 + \ldots ,
	\label{eq:UGellMann}
\end{equation}
where $\lambda_i$, $i=1,\ldots,8$, are the usual Gell-Mann matrices,
$\lambda_9$ is the unit matrix, and the $\alpha_i^{(j)}$ are real parameters that are to be 
determined.\footnote{Note that we cannot assume that the $U^{(j)}$ are special
unitary, which is why we need to include a ninth matrix in our basis.}  
The idea of the procedure is to determine matrices $U^{(1)}$, $U^{(2)}$, $U^{(3)}$, \ldots, in such a 
way that $U^{(m)}$ approaches the identity matrix for large $m$ (i.e., in such a way that the $\alpha_i^{(m)}$
approach zero for large $m$).  This allows one to truncate the infinite product
in Eq.~(\ref{eq:Ue-product}), so that $U_e^{(m_\textrm{\tiny max})}$ (for some $m_\textrm{\scriptsize  max}$) 
is used as a suitable approximation to $U_e$.

In the first step of the procedure we substitute 
\begin{equation}
	U_e^{(1)} = U^{(1)} \tilde{I}
    \label{eq:Ue1}
\end{equation}
into Eq.~(\ref{eq:Ueme}), in place of $U_e$, and 
then expand the expression for $U^{(1)}$ (see Eq.~(\ref{eq:UGellMann})) to linear
order in the $\alpha_i^{(1)}$.  This gives us the defining expression for the nine unknowns $\alpha_i^{(1)}$,
\begin{equation}
	\left(1 + \sum_{i=1}^9\frac{i \alpha_i^{(1)}}{2} \lambda_i \right) \tilde{I} m_e - 
        m_e \tilde{I} \left(1 - \sum_{i=1}^9\frac{i \alpha_i^{(1)}}{2} \lambda_i \right) = i s_a t_{2\beta}{\cal M} \; .
    \label{eq:alphai1-original}
\end{equation}
Defining $\widetilde{m}_e^{(0)} \equiv \tilde{I} m_e$ and rearranging, we have
\begin{equation}
	\sum_{i=1}^9\frac{i \alpha_i^{(1)}}{2} \left(\lambda_i \widetilde{m}_e^{(0)} + \widetilde{m}_e^{(0)\dagger} \lambda_i \right)
        = i s_a t_{2\beta}{\cal M} - \widetilde{m}_e^{(0)} + \widetilde{m}_e^{(0)\dagger} \; .
\end{equation}
To solve for the $\alpha_i^{(1)}$, we multiply the above expression by $\lambda_k$ and take the trace,
\begin{equation}
	\sum_{i=1}^9\frac{i \alpha_i^{(1)}}{2} \mbox{Tr}\left[\left(\lambda_i \widetilde{m}_e^{(0)} 
        + \widetilde{m}_e^{(0)\dagger} \lambda_i \right) \lambda_k \right]
        = \mbox{Tr}\left[\left(i s_a t_{2\beta}{\cal M} - \widetilde{m}_e^{(0)} + \widetilde{m}_e^{(0)\dagger} \right) \lambda_k \right] \; ,
    \label{eq:alphai1}
\end{equation}
which yields nine equations (for $k=1,2,\ldots, 9)$ in the nine unknowns.  
Once we have determined the $\alpha_i^{(1)}$, we use the 
series expansion of Eq.~(\ref{eq:UGellMann}) to determine the matrix $U^{(1)}$ (summing up enough
terms so that the result is very close to unitary).  While the linearized version of $U^{(1)}$
was an exact solution of Eq.~(\ref{eq:alphai1-original}), the ``re-unitarized'' version of the matrix (i.e., $U^{(1)}$)
is no longer a solution when Eq.~(\ref{eq:Ue1}) is substituted into Eq.~(\ref{eq:Ueme}).  This brings
us to the next step in the procedure.

In the second step we substitute 
\begin{equation}
	U_e^{(2)} = U^{(2)} U^{(1)} \tilde{I},
\end{equation}
into Eq.~(\ref{eq:Ueme}) and expand $U^{(2)}$ to linear order in the coefficients $\alpha_i^{(2)}$,
while keeping the matrix $U^{(1)}$ in its exactly unitary form.
Performing the same manipulations as in the previous step, we obtain
the following expression for the $\alpha_i^{(2)}$,
\begin{equation}
	\sum_{i=1}^9\frac{i \alpha_i^{(2)}}{2} \mbox{Tr}\left[\left(\lambda_i \widetilde{m}_e^{(1)} 
        + \widetilde{m}_e^{(1)\dagger} \lambda_i \right) \lambda_k \right]
        = \mbox{Tr}\left[\left(i s_a t_{2\beta}{\cal M} - \widetilde{m}_e^{(1)} + \widetilde{m}_e^{(1)\dagger} \right) \lambda_k \right] \; ,
    \label{eq:alphai2}
\end{equation}
where $\widetilde{m}_e^{(1)} \equiv U^{(1)} \tilde{I} m_e = U^{(1)} \widetilde{m}_e^{(0)}$.
This allows us to solve for the $\alpha_i^{(2)}$ and to exponentiate the corresponding sum to determine
$U^{(2)}$.

We continue in this manner, linearizing $U^{(m)}$ at the $m$th step in order to determine
the coefficients $\alpha_i^{(m)}$ (while using 
the exactly unitary versions of the matrices determined in the previous steps) and then ``re-unitarizing''
at the end to obtain $U^{(m)}$.  After several iterations, the coefficients become vanishingly small
and we terminate the process, having obtained
an approximation to $U_e$ that is unitary and satisfies Eq.~(\ref{eq:Ueme}) to a high degree of accuracy.

We conclude with two comments:
\begin{enumerate}
\item The matrix ${\cal M}$ that is produced in Step~\ref{step:calM} of the iterative process 
described in Sec.~\ref{subsec:iterative-method} is actually only an approximation
to the exact expression and may not be exactly Hermitian.  In practice, therefore, we replace ${\cal M}$ 
by $\frac{1}{2}\!\left({\cal M} + {\cal M}^\dagger\right)$ wherever it appears in the expressions in this
Appendix.\footnote{Even after coercing ${\cal M}$ into a Hermitian form it is not guaranteed
that a unitary matrix $U_e$ exists such that Eq.~(\ref{eq:Ueme}) is satisfied.  To see that this is the case
we only need consider the $1\times 1$ case, in which $U_e$ is a pure phase and ${\cal M}$ is a real number.
For ${\cal M}$ larger than a certain value there is no longer a solution for $U_e$.  Our method implicitly
assumes that the expressions produced for ${\cal M}$ are sufficiently close to the ``true''
expression that a solution exists for $U_e$.}

\item Equations~(\ref{eq:alphai1}) and (\ref{eq:alphai2}) and
the analogous expressions for the subsequent steps in the process guarantee 
that the $\alpha_i^{(j)}$ will be real 
if ${\cal M}$ is Hermitian and if unique solutions exist.
When the $\alpha_i^{(j)}$ are determined numerically they generically include small
imaginary parts.  We discard these.
\end{enumerate}

\section{Angular parametrization of $SO(3,\mathbb{C})$}
\label{sec:angular-parametrization}

In this appendix, we show that almost all $3\times 3$ complex orthogonal matrices of determinant $1$ can be realized as 
$$
\left(
\begin{array}{ccc}
c_{\eta_1} c_{\eta_3} -c_{\eta_2} s_{\eta_1} s_{\eta_3}  & s_{\eta_1} s_{\eta_2}  & c_{\eta_1} s_{\eta_3} +c_{\eta_2} c_{\eta_3} s_{\eta_1}  \\ 
s_{\eta_2} s_{\eta_3}  & c_{\eta_2}  & -c_{\eta_3}  s_{\eta_2}  \\ 
-c_{\eta_3} s_{\eta_1} -c_{\eta_1} c_{\eta_2} s_{\eta_3}  & c_{\eta_1} s_{\eta_2}  & c_{\eta_1} c_{\eta_2} c_{\eta_3}  - s_{\eta_1} s_{\eta_3} 
\end{array}
\right),
$$
where $s_{\eta_i}=\sin(\eta_i)$ and $c_{\eta_i}=\cos(\eta_i)$,
for suitable choices of complex angles $\eta_i\in \mathbb{C}$. This generalizes the well-known parametrization of $SO(3,\mathbb{R})$ using Euler angles (see for example Ref.~\cite{Biedenharn1981}). 

We start with a lemma that we will need later. 

\begin{lemma}\label{ang}
Suppose $v^2+w^2=1$ for complex numbers $v,w$. Then there exists a complex angle $\eta$ such that $v = \cos \eta, w = \sin\eta$. 
\end{lemma}

Note that $(v+iw)(v-iw) = 1$, so $v+iw\ne 0$. Find any complex angle $\eta$ such that $v+iw = e^{i\eta}$. This is possible since $v+iw\ne 0$, and the range of $e^z$ is all nonzero complex numbers. 

Then $e^{-i\eta} = 1/(v+iw) = v-iw$. Therefore 
$$
\cos\eta = \frac{e^{i\eta}+e^{-i\eta}}{2} = \frac{2v}{2} = v
$$
and 
\begin{align*}
\sin\eta = \frac{e^{i\eta}-e^{-i\eta}}{2i} =\frac{2iw}{2i} &= w.  
\end{align*}

\subsection{Recovering a $3\times 3$ special orthogonal matrix from the middle cross}

\begin{proposition}\label{recover}
Let 
$$
\left(
\begin{array}{ccc}
a & p & x \\
b & q & y \\
c & r & z
\end{array}
\right)
$$
be an arbitrary element of $SO(3,\mathbb{C})$. Suppose $q^2\ne 1$. Then 
\begin{align*}
a &= \frac{-ry-bqp}{1-q^2}; & x &= \frac{br-qpy}{1-q^2};\\
c &= \frac{py-bqr}{1-q^2}; & 
z &= \frac{-bp-qry}{1-q^2}.
\end{align*}
\end{proposition}

If $p$ and $r$ were both equal to $0$, then the normalization of the second column would imply $q^2=1$, which cannot be. So we may assume $p\ne 0$ (up to symmetry). Likewise, we will assume $y\ne 0$, following a similar argument for the second row. 

From the orthogonality of the first two columns, i.e., $ap+bq+cr=0$, we get 
\begin{align}\label{a-eventual}
ap=-bq-cr,
\end{align}
which we may substitute in the normalization condition for the first column, namely $a^2p^2+b^2p^2+c^2p^2=p^2$, to obtain
$$
b^2q^2+c^2r^2+2bcqr + b^2 p^2 + c^2 p^2 = p^2,
$$
which is a quadratic equation in $c$:
$$
(1-q^2)c^2 + (2bqr)c + b^2(q^2+p^2)-p^2 = 0 .
$$
The discriminant of this equation is 
\begin{align*}
4b^2q^2r^2-4(1-q^2)\left(b^2(q^2+p^2)-p^2\right) &= 4\left(b^2q^2r^2-(1-q^2)(b^2-b^2r^2-p^2)   \right)\\
&= 4\left( -b^2 +b^2r^2+p^2  +q^2 b^2 - q^2 p^2  \right)\\
&= 4\left( b^2(r^2+q^2-1) + p^2(1-q^2)   \right)\\
&= 4\left( -b^2 p^2 + p^2 (1-q^2) \right)\\
&= 4\left(p^2 y^2\right)\\
&=(2py)^2.
\end{align*}
Thus 
\begin{align}\label{c-form}
c = \frac{-2bqr + \epsilon\cdot2py}{2(1-q^2)} = \frac{-bqr + \epsilon py}{1-q^2},
\end{align}
where $\epsilon$ is either $1$ or $-1$: to be determined.

Since $p\ne0$, we may divide Eq. (\ref{a-eventual}) by $p$ and substitute Eq. (\ref{c-form}) for $c$ to obtain
$$
a = \frac{bqr^2-\epsilon pyr}{p(1-q^2)}-\frac{bq}{p} = \frac{bqr^2-\epsilon pyr - bq (p^2+r^2)}{p(1-q^2)} = \frac{-\epsilon yr - bqp}{1-q^2}.
$$

In similar fashion, since $y\ne 0$, we obtain 
$$
x = \frac{-pqy+\epsilon b r}{1-q^2}
$$
and 
$$
z = \frac{-q r y - \epsilon b p}{1-q^2}.
$$

By a straightforward calculation, one finds that 
$$
\operatorname{det} 
\left(
\begin{array}{ccc}
\frac{-\epsilon yr - bqp}{1-q^2} & p & \frac{-pqy+\epsilon b r}{1-q^2} \\
b & q & y \\
\frac{-bqr + \epsilon py}{1-q^2} & r & \frac{-q r y - \epsilon b p}{1-q^2}
\end{array}
\right)
$$
reduces simply to $\epsilon$. Therefore $\epsilon=1$ and we are done.

\subsection{Complex Euler angles}

\begin{theorem}\label{euler-angles}
Let 
$$A = \left(
\begin{array}{ccc}
a & p & x \\
b & q & y \\
c & r & z
\end{array}
\right)
$$
be an element of $SO(3,\mathbb{C})$ such that $q^2\ne1$. Then there exist complex angles $\eta_1,\eta_2,\eta_3$ such that 
$$
A = 
\left(
\begin{array}{ccc}
c_{\eta_1}c_{\eta_3}-c_{\eta_2}s_{\eta_1}s_{\eta_3} & s_{\eta_1}s_{\eta_2} & c_{\eta_1}s_{\eta_3}+c_{\eta_2}c_{\eta_3}s_{\eta_1} \\ 
s_{\eta_2}s_{\eta_3} & c_{\eta_2} & -c_{\eta_3} s_{\eta_2} \\ 
-c_{\eta_3}s_{\eta_1}-c_{\eta_1}c_{\eta_2}s_{\eta_3} & c_{\eta_1}s_{\eta_2} & c_{\eta_1}c_{\eta_2}c_{\eta_3} - s_{\eta_1}s_{\eta_3}
\end{array}
\right)
$$
$($where $c_{\eta_i} = \cos(\eta_i)$ and $s_{\eta_i} = \sin(\eta_i))$. 
\end{theorem}

Find a complex number $u$ such that $u^2 = 1-q^2$. By Lemma \ref{ang}, there is a complex angle $\eta_2$ such that $\cos(\eta_2) = q, \sin(\eta_2) = u$. Note that $u\ne 0$; moreover 
$$
\left(\frac{p}{u}\right)^2 + \left(\frac{r}{u}\right)^2 = \frac{p^2+r^2}{1-q^2} = 1,
$$
so there exists a complex angle $\eta_1$ such that $s_{\eta_1} = p/u$ and $c_{\eta_1} = r/u$. Thus $p = s_{\eta_1}s_{\eta_2}$ and $r = c_{\eta_1}s_{\eta_2}$. 

Similarly, there exists a complex angle $\eta_3$ such that $b = s_{\eta_2}s_{\eta_3}$ and $-y = c_{\eta_3}s_{\eta_2}$ (here we apply Lemma \ref{ang} to the pair $b/u$, $-y/u$.)

Now we apply Proposition \ref{recover}: 
$$
a = \frac{-ry-bqp}{u^2} = \frac{c_{\eta_1}c_{\eta_3}s_{\eta_2}^2 - s_{\eta_3}s_{\eta_1}c_{\eta_2}s_{\eta_2}^2}{s_{\eta_2}^2} = c_{\eta_1}c_{\eta_3} - c_{\eta_2}s_{\eta_1}s_{\eta_3}.
$$
Similarly, 
\begin{align*}
c &= -c_{\eta_3}s_{\eta_1} - c_{\eta_1}c_{\eta_2}s_{\eta_3} \\
x &= c_{\eta_1}s_{\eta_3}+c_{\eta_2}c_{\eta_3}s_{\eta_1} \\
z &= c_{\eta_1}c_{\eta_2}c_{\eta_3} - s_{\eta_1}s_{\eta_3}. \qedhere
\end{align*}

\subsection{Other cases}

If it happens that $q^2=1$ in the above orthogonal matrix $A$, then there are other ``Euler angles'' that may be used to parametrize its entries. The easiest way to obtain this is to replace $A$ by $A'=PAQ$ for suitable permutation matrices $P$ and $Q$, such that the $(2,2)$ entry of $A'$ is not $\pm1$. In other words, even if $q^2=1$, there must be some other entry of $A$ that does not square to $1$. Permute rows and columns of $A$ until that entry is now in the $(2,2)$ position, and then apply Theorem \ref{euler-angles}.

\bibliographystyle{JHEP}

\bibliography{paper}

\end{document}